\documentclass[twocolumn,10pt]{asme2ej}

\usepackage{ntheorem}
\usepackage{graphicx}
\usepackage{float}
\usepackage{epsfig} % for postscript graphics files
\usepackage{newtxmath} % necessary for \boldsymbol{}
\usepackage{bbm} % used for ones vector 
\usepackage{times} % assumes new font selection scheme installed
\usepackage{amsmath} 

\usepackage{color}
\usepackage{mathrsfs}
\usepackage{hyperref}

\def\bs{\boldsymbol}
\def\inr{\in\mathbb{R}}
\newcommand{\norm}[1]{\left\lVert#1\right\rVert}

\makeatletter
\let\NAT@parse\undefined
\makeatother

\usepackage[square, sort&compress, numbers]{natbib}

\def\real{\mathbb{R}}

\newcommand{\until}[1]{\{1,\dots, #1\}}

\newcommand{\supscr}[2]{#1^{\textup{#2}}}
\newcommand{\setdef}[2]{\{#1 \; | \; #2\}}

\newcommand\oprocendsymbol{\hbox{$\square$}}
\newcommand\oprocend{\relax\ifmmode\else\unskip\hfill\fi\oprocendsymbol}

\newtheorem{theorem}{Theorem}

\newtheorem{lemma}{Lemma}

\newtheorem{remark}{Remark}

\newtheorem{assumption}{Assumption}
\newtheorem{definition}{Definition}

\renewcommand{\theenumi}{\roman{enumi}}

\title{A High-Gain Observer Approach to Robust Trajectory Estimation and Tracking for a Multi-rotor UAV \thanks{This work has been supported in part by NASA-MSGC award NNX15AJ20H and NSF grant IIS-1734272. An earlier version of this work \cite{boss2020ACC} was presented at the 2020 American Control Conference. In addition to the ideas presented in \cite{boss2020ACC}, this paper contains rigorous mathematical analysis as well as an experimental implementation of the proposed control strategy.}}

\author{Connor J. Boss 
	\affiliation{
	Department of Electrical and Computer Engineering\\
		Michigan State University\\
		 East Lansing, MI, 48824 \\
		Email: connorjboss@gmail.com}}
\author{Vaibhav Srivastava 
	\affiliation{
		Department of Electrical and Computer Engineering\\
		Michigan State University\\
		East Lansing, MI, 48824 \\
		Email: vaibhav@egr.msu.edu
	}	
}
\begin{document}

\maketitle 

\begin{abstract}
\emph{Using the context of trajectory estimation and tracking for multi-rotor unmanned aerial vehicles (UAVs), we explore the challenges in applying high-gain observers to highly dynamic systems. The multi-rotor will operate in the presence of external disturbances and modeling errors. At the same time, the reference trajectory is unknown and generated from a reference system with unknown or partially known dynamics. We assume the only measurements that are available are the position and orientation of the multi-rotor and the position of the reference system. We adopt an extended high-gain observer (EHGO) estimation framework to estimate the unmeasured multi-rotor states, modeling errors, external disturbances, and the reference trajectory. We design a robust output feedback controller for trajectory tracking that comprises a feedback linearizing controller and the EHGO. The proposed control method is rigorously analyzed to establish its stability properties. Finally, we illustrate our theoretical results through numerical simulation and experimental validation in which a multi-rotor tracks a moving ground vehicle with an unknown trajectory and dynamics and successfully lands on the vehicle while in motion.}
\end{abstract}

\section{Introduction}
Extended high-gain observers~\cite{khalil2017HGO,astolfi2015high} have been applied to many physical systems, from electro-hydraulic actuators~\cite{tran2019extended} to permanent magnet synchronous motors~\cite{khalil2017high}, robotic manipulators~\cite{lee1997adaptive,yang2010robust}, cart-pendulum system~\cite{lee2015output}, and aerial vehicles~\cite{lee2021output, lee2012control,marconi2014high}. However, due to noise amplification~\cite{ahrens2009high,ball2011analysis} and high sample-rate requirements, EHGOs are typically not applied to highly dynamic systems. For these highly dynamic systems, the required timescale of the EHGO can become fast enough such that it matches the timescale of the subsystem (e.g., actuator) dynamics which are typically ignored in the control design. In this case, the fast subsystem dynamics must also be incorporated into the observer design, and if omitted can induce instabilities in the system. Exploring the application of EHGOs to multi-rotor control to increase robustness through the estimation of uncertainties and disturbances, we address these challenges and show that EHGOs can indeed be implemented on highly dynamic systems at reasonable sample rates with inexpensive sensors.

In this paper, we focus on implementing EHGOs on highly dynamic systems using the context of multi-rotor drones. We choose a multi-rotor drone as our example system because of its broadening applications. Much work has been devoted to studying multi-rotor UAV control design; see~\cite{papachristos2018modeling,kumar2012opportunities} for a survey. Numerous linear and nonlinear control approaches have been applied to multi-rotors, including PID~\cite{huang2009aerodynamics}, feedback linearization~\cite{voos2009nonlinear,lee2009feedback}, adaptive backstepping control~\cite{huang2010adaptive,lee2011adaptive}, and model predictive control (MPC)~\cite{nguyen2021model}to name a few. Linear methods are effective near the hover configuration but can experience degraded performance during aggressive maneuvers. Feedback linearization is sensitive to sensor noise as well as model uncertainty~\cite{lee2009feedback}, but results in a linear system that is simple to analyze. Adaptive backstepping  control design~\cite{huang2010adaptive,lee2011adaptive} enables unknown system parameters to be estimated during operation to reduce the effect of model uncertainty. MPC enables excellent tracking performance and can achieve aggressive maneuvers~\cite{nguyen2021model}, but at the cost of computational complexity. 

Additionally, many linear and nonlinear approaches are subject to reduced performance in the presence of model uncertainty and external disturbances~\cite{thanh2018quadcopter}. This motivates the use of robust methods which can overcome certain classes of disturbances utilizing observers~\cite{thanh2018quadcopter,kim2018robust} or adaptive approaches to estimate model parameters online~\cite{lee2011adaptive}. The magnitude of disturbance that can be canceled may depend on the control gains, which can be tuned using an adaptive gain scheduling approach~\cite{thanh2018quadcopter}.

In contrast to these approaches, we adopt a disturbance observer approach that allows handling (i) partial state measurement, (ii) imperfectly known system parameters, and (iii) unknown disturbances. 
Furthermore, the magnitude of disturbance EHGO-based controllers can overcome is not dependent on control gains, but rather on observer gains. This allows total freedom in assigning control gains, which can be chosen to shape the transient response as we recover the performance of a desired linear system. Additionally, our EHGO-based approach also allows for estimating a desired system trajectory in the contexts where the desired trajectory is generated by a dynamical system with unknown or partially known dynamics. Examples include trajectories generated by a human-driven vehicle or a boat operating in unsteady waters. 

To showcase the performance of our method, we will apply our technique to the problem of landing a multi-rotor on a mobile platform~\cite{kong2014vision,gautam2014survey}. Multiple control methodologies have been applied to this problem, including model predictive control~\cite{feng2018autonomous,maces2017autonomous}, PI control \cite{herisse2011landing,serra2016landing}, and feedback linearizing control~\cite{hoang2017vision}. Many approaches either do not consider modeling error and external disturbances or consider them to be constant or slowly time-varying~\cite{herisse2011landing,serra2016landing}. In contrast, our approach only requires that any uncertainty be bounded and continuously differentiable. 

While other methods may only address a subset of these challenges, EHGOs can provide a unified framework that can ensure flight performance under model uncertainties, external disturbances, and unknown reference trajectory dynamics. It can enable the design of a robust feedback linearizing control strategy that can achieve excellent transient tracking performance even when applied to this highly dynamic system in the presence of these disturbances. It also enables real-time trajectory generation based on the position information of a reference system which may have partially known or completely unknown dynamics. However, effectively implementing EHGOs with limited measurement sampling rate is challenging for highly dynamic systems and we address this challenge in this paper. 

In this work, we design and rigorously analyze an EHGO-based feedback linearizing control method that incorporates the estimation of a reference trajectory from an unknown, or partially known, reference dynamic system. While being implemented on a highly dynamic system, the EHGO enables the estimation of all states for output feedback control, as well as estimating modeling error and external disturbances, enabling the design of a robust feedback linearizing control strategy. The proposed method can recover the performance of the desired linear system under a broad class of disturbances. 
 
The major contributions of this work are threefold. First, our proposed EHGO-based control design for the translational system treats the transient of the rotational subsystem as a disturbance, estimates it, and actively compensates for it. This enables the proposed controller to not require the timescale separation between rotational and translational subsystems, a key requirement in existing multi-rotor control techniques. 

Second, we show that for the output sampling rates available on the off-the-shelf drones, the dynamics of commercial electronic speed controllers evolve at the same timescale as the EHGO. We illustrate the influence of inclusion/non-inclusion of these dynamics on output feedback performance and show that these should be included in the EHGO design. Additionally, the output feedback control dynamics obtained after 
the inclusion of these dynamics differs from the standard EHGO-based  output feedback dynamics, and we rigorously analyze them. 

Third, we illustrate the effectiveness of our output feedback controller through simulation and experimental results using the example of landing a multi-rotor on a moving ground vehicle and confirm that EHGOs can be effectively implemented in this class of highly dynamic systems. Our implementation involves a hexrotor whose size is roughly 1/6-th of the flying arena and this leads to large aerodynamic effects due to the interaction of the rotors' airflow with physical structures. We show that the EHGO (i) estimates disturbances due to these aerodynamic effects and allows the controller to compensate for them, (ii) estimates the trajectory of the ground vehicle, and (iii) enables autonomous landing of the hexrotor on a ground vehicle smaller than its size. 

The remainder of the paper is organized as follows. The system dynamics are introduced in Section \ref{sec:systemDynamics} with the control and observer design in Section \ref{sec:ControlDesign}. The controller is analyzed in Section \ref{sec:stabilityAnalysis} and is validated through simulation in Section \ref{sec:simulation} with experimental results presented in Section \ref{sec:experiment}. Conclusions are presented in Section \ref{sec:conclusion}. Detailed proofs of the technical results are provided in the appendix.

\begin{table}[t!]\label{tab:listOfSymbols}
\centering
\caption{List of Symbols}
\resizebox{0.95\linewidth}{!}{
\begin{tabular}{cl}
\hline
$\bs{\tau} \inr^3$ \quad & Torque applied to rigid body\\
$J\inr^{3\times3}$ \quad &Inertia matrix\\
$n \in \mathbb N$ &Number of rotors \\
$\Omega\inr^3$ \quad & Angular velocity of rigid body\\
$\bs{\vartheta}_1 = [\phi \ \theta \ \psi]^\top \in \mathbb T^3$ & ZYX Euler angles\\
$\bs{\vartheta}_2 = [\dot{\phi} \ \dot{\theta} \ \dot{\psi}]^\top \inr^3$ &Time derivative of $\bs{\vartheta}_1$\\
$\bs{\sigma}_\star \inr^3$  &Disturbance term in $\star$ dynamics \\
$\bs{\varsigma}_\xi \inr^3$ & Rotational disturbance term \\
$\bs{\xi}_1 \inr^3$  &Rotational tracking error \\
$\bs{\xi}_2 \inr^3$ &Angular rate tracking error \\
$\star_r \inr^3$  &Reference signal for $\star$ system \\
$\dot{\bar{\bs{\vartheta}}}_r \inr^3$ &Approximation of reference signal $\dot{\bs{\vartheta}}_r$ \\
$\bs{e}_z = [0 \ 0 \ 1]^\top \inr^3$ &Unit vector in $z$ direction \\
$R_3(\bs{\vartheta}_1)$ &Third column of rotation matrix \\
$\bs{p}_1 \inr^3$ &Position in inertial frame \\
$\bs{p}_2 \inr^3$  &Velocity in inertial frame \\
$\bs{\rho}_1 \inr^3$  &Position error \\
$\bs{\rho}_2 \inr^3$ &Velocity error \\
$v_f \inr$ \quad &Collective thrust control input \\
$\bs{p}_{c_1} \inr^3$  &Reference system position \\
$\bs{p}_{c_2} \inr^3$  &Reference system velocity \\
$C_T \inr_{>0}$ &Rotor thrust coefficient \\
$C_D \inr_{>0}$  &Rotor aerodynamic drag coefficient \\
$\tau_m \inr_{>0}$ &Actuator time constant \\
$\bar{f}_i \inr_{>0}$ &Thrust generated by $i$-th rotor \\
$\omega_i \inr_{>0}$ &$i$-th rotor angular rate \\
$\bs{\omega}_s \inr^n$  &Vector of $\omega_i^2$ \\
$\bs{\omega}_{des} \inr^n$ &Desired rotor speeds \\
$M \inr^{4\times n}$ &Allocation matrix \\
$\bs{e}_\vartheta \inr^3$  &Perturbation due to rotational tracking error \\
$\bs{u}_r \inr^3$  &Virtual rotational control input \\ 
$\beta_1, \ \beta_2 \inr_{>0}$ \quad &Rotational control gains \\
$\bs{u}_t \inr^3$  &Virtual translational control input \\
$\gamma_1, \ \gamma_2 \inr_{>0}$ \quad & Translational control gains \\
$\bs{\chi} \inr^{30}$  & Full state vector of EHGO \\
$\alpha_1, \ \alpha_2, \ \alpha_3, \ \alpha_4 \inr_{>0}$  & Observer gains \\
$\epsilon \inr_{>0}$  & High-gain parameter of observer \\
$\hat{\star}$   &Estimate of $\star$ from EHGO \\
$\hat{\bs{\tau}}_d \inr^3$ &Output feedback desired torque\\
$\hat{v}_{f d} \inr$ &Output feedback desired thrust\\
$V_\star$ & Lyapunov function for $\star$ dynamics \\
$\Omega_\star$ & Region of attraction for $\star$ dynamics \\
$c_\star$  & Bound on Lyapunov function, $V_\star$ \\
$\bs{\eta} \inr^{30}$ & Observer estimation errors 
\\
\hline
\end{tabular}}
\end{table}

\section{System Dynamics}\label{sec:systemDynamics}
In this section, we review the dynamics of the different subsystems of a multi-rotor UAV and reference system. 

\subsection{Rotational Dynamics}
The rotational dynamics of the multi-rotor are 
\begin{equation}\label{eq:rotational-dynamics}
    \bs{\tau} = J\dot{\Omega} + \Omega \times J\Omega,
\end{equation}
where $J\inr^{3\times3}$ is the inertia matrix, $\bs{\tau}\inr^3$ is the torque applied to the multi-rotor body and $\Omega\inr^3$ is the angular velocity, each expressed in the body-fixed frame~\cite{lee2010geometric}. A full list of symbols can be found in Table \ref{tab:listOfSymbols}.

Consider the orientation of the multi-rotor expressed in terms of Z-Y-X Euler angles $\bs{\vartheta}_1 = [\phi \ \theta \ \psi]^\top\in (-\frac{\pi}{2},\frac{\pi}{2})^2 \times \ (-\pi,\pi]$. The angular velocity, $\Omega$, is related to the Euler angle rates $\bs{\vartheta}_2 = [\dot{\phi} \ \dot{\theta} \ \dot{\psi}]^\top \inr^3$ in the inertial frame as

\begin{equation}
\begin{split}
    \bs{\vartheta}_2 &= \Psi\Omega, \quad \Psi = \left[\begin{matrix}
    1&\sin(\phi) \tan(\theta)&\cos(\phi) \tan(\theta)\\
    0&\cos(\phi)&-\sin(\phi)\\
    0&\sin(\phi)/\cos(\theta)&\cos(\phi)/\cos(\theta) \end{matrix}\right],\\
    \Omega &= \Psi^{-1}\bs{\vartheta}_2,
\end{split}
\end{equation}

The rotational dynamics can be equivalently written in terms of Euler angles as
\begin{equation}\label{eq:rotationalDynamics}
    \begin{split}
        \dot{\bs{\vartheta}}_1 &= \bs{\vartheta}_2\\
        \dot{\bs{\vartheta}}_2 &= \dot{\Psi} \Psi^{-1} \bs{\vartheta}_2 - \Psi J^{-1}(\Psi^{-1} \bs{\vartheta}_2 \times J\Psi^{-1} \bs{\vartheta}_2)\\
        & \quad + \Psi J^{-1}\bs{\tau} + \bs{\sigma}_\xi,\\
    \end{split}
\end{equation}
where $\bs{\sigma}_\xi \inr^3$ is an added term to represent the lumped rotational disturbance and satisfies \textit{Assumption \ref{as:disturbance}} below.
\begin{definition}[Prime Canonical Form]\label{def:primeCanonical}
    A control system in the ``prime canonical form''~\cite{morse1973structural}, for state $\bs{x} \in \real^n$, control input $u \inr$, and disturbance $\sigma \inr$, has the following representation
    \begin{equation}\label{eq:pcf}
        \dot{\bs{x}} = A_{\textup{prm}} \bs{x} + B_{\textup{prm}} f_{\textup{prm}}(t,\bs{x},u), \quad y = C_{\textup{prm}} \bs{x},
    \end{equation}
    where 
    \begin{equation}
        \begin{split}
        A_{\textup{prm}} = \left[\begin{smallmatrix} 0_{n-1 \times 1} & I_{n-1} \\ 0 & 0_{1\times n-1} \end{smallmatrix}\right], \; \; B_{\textup{prm}}= \left[\begin{smallmatrix} 0_{n-1 \times 1} \\ 1 \end{smallmatrix}\right], \; \; \\
        f_{\textup{prm}}: \real_{\ge 0} \times \real^n \times \real \to \real, \; \; C_{\textup{prm}} = [1 \; 0_{1 \times n-1}],
        \end{split}
    \end{equation}
    $0_{p\times q}$ is a matrix of zeros with dimension $p\times q$, $I_p$ is the identity matrix of dimension $p$, and $y\inr$ is the measurement.
\end{definition}
\begin{assumption}[Disturbance Properties]\label{as:disturbance}
The dynamics of the various subsystems in this work take the prime canonical form perturbed by a disturbance term. We assume that the disturbance enters the RHS of \eqref{eq:pcf} as $B_d \sigma$ where $B_d = B_{\textup{prm}}$, $\sigma$ is continuously differentiable, and its partial derivatives with respect to states are bounded on compact sets of those states for all $t\geq0$.
\end{assumption}

Let $\bs{\vartheta}_r = [\phi_r \ \theta_r \ \psi_r]^\top \in (-\frac{\pi}{2},\frac{\pi}{2})^2 \times (-\pi,\pi]$ and $\dot{\bs{\vartheta}}_r = [\dot{\phi}_r \ \dot{\theta}_r \ \dot{\psi}_r]^\top \inr^3$ be the rotational reference signals. Define the rotational tracking error variables
\begin{equation}
    \begin{gathered}
        \bs{\xi}_1 = \bs{\vartheta}_1 - \bs{\vartheta}_r, \quad \bs \xi_2 = \dot{\bs{\xi}_1} = \bs{\vartheta}_2 - \dot{\bs{\vartheta}}_r, \quad \bs \xi = [\bs \xi_1^\top \; \bs \xi_2^\top]^\top.
    \end{gathered}
\end{equation}
The rotational dynamics \eqref{eq:rotationalDynamics} can now be written in terms of tracking error
\begin{equation}\label{eq:rotationalErrorDynamics}
    \begin{split}
        \dot{\bs{\xi}}_1 &= \bs{\xi}_2 \\
        \dot{\bs{\xi}}_2 &= f(\bs{\xi},\bs{\vartheta}_1,\dot{\bs{\vartheta}}_r) + G(\bs{\vartheta}_1)\bs{\tau} + \bs{\sigma}_\xi - \ddot{\bs{\vartheta}}_r,
    \end{split}
\end{equation}
where
\begin{equation}
    \begin{split}
        f(\bs{\xi},\bs{\vartheta}_1,\dot{\bs{\vartheta}}_r) &= \dot{\Psi}\Psi^{-1}(\bs{\xi}_2 + \dot{\bs{\vartheta}}_r) \\
        &\quad -\Psi J^{-1}(\Psi^{-1}(\bs{\xi}_2 + \dot{\bs{\vartheta}}_r) \times J\Psi^{-1}(\bs{\xi}_2 + \dot{\bs{\vartheta}}_r)) \\
        G(\bs{\vartheta}_1) &= \Psi J^{-1}.
    \end{split}
\end{equation}
Suppose that only $\dot{\bar{\bs{\vartheta}}}_r$, an approximation of $\dot{\bs{\vartheta}}_r$, is known. Then \eqref{eq:rotationalErrorDynamics} can be rewritten as
\begin{equation}\label{eq:rotationalErrorDynamicsFinal}
    \begin{split}
        \dot{\bs{\xi}}_1 &= \bs{\xi}_2 \\
        \dot{\bs{\xi}}_2 &= f(\bs{\xi},\bs{\vartheta}_1,\dot{\bar{\bs{\vartheta}}}_r) + G(\bs{\vartheta}_1)\bs{\tau} + \bs{\varsigma}_\xi,
    \end{split}
\end{equation}
where $\bs{\varsigma}_\xi = \bs{\sigma}_\xi - \ddot{\bs{\vartheta}}_r + [f(\bs{\xi},\bs{\vartheta}_1,\dot{\bs{\vartheta}}_r) - f(\bs{\xi},\bs{\vartheta}_1,\dot{\bar{\bs{\vartheta}}}_r)]$, which also satisfies \textit{Assumption \ref{as:disturbance}} based on the properties of $f$ and by assuming the reference trajectory is third-order continuously differentiable.

\begin{remark}\label{rem:referenceApproximation}
Note that $\dot{\bar{\bs{\vartheta}}}_r$ is an approximation of $\dot{\bs{\vartheta}}_r$ which should be chosen appropriately based on the available information. The choice of $\dot{\bar{\bs{\vartheta}}}_r$ will determine how big the difference term, $f(\bs{\xi},\bs{\vartheta}_1,\dot{\bs{\vartheta}}_r) - f(\bs{\xi},\bs{\vartheta}_1,\dot{\bar{\bs{\vartheta}}}_r)$, in $\bs{\varsigma}_\xi$ will be. This will in turn determine the magnitude of the rotational disturbance term $\bs{\varsigma}_\xi$. If for example we assume we have no information of how $\dot{\bs{\vartheta}}_r$ behaves, we can set $\dot{\bar{\bs{\vartheta}}}_r = 0$. This will lead to an increase in the magnitude of $\bs{\varsigma}_\xi$ because it will contain this large difference between $\dot{\bar{\bs{\vartheta}}}_r$ and $\dot{\bs{\vartheta}}_r$. As a result, the disturbance estimator will need to work harder to estimate the larger magnitude disturbance and may require increasing the gain of the observer to achieve desired performance. As a result, any information that can be used to partially reconstruct $\dot{\bs{\vartheta}}_r$ should be used to reduce the burden on the observer.
\end{remark}

\subsection{Translational Dynamics}
Let $\bs{p}_1 = [x \ y \ z]^\top\inr^3$ and $\bs{p}_2 = [\dot{x} \ \dot{y} \ \dot{z}]^\top\inr^3$, respectively, be the position and velocity of the multi-rotor center of mass expressed in the inertial frame. Let the thrust generated by the $i$-th rotor be $\bar{f}_i \inr$, and the total thrust force, $v_f = \sum_{i=1}^n{\bar{f}_i} \inr$ serves as the input to the translational system. Let the mass of the aerial platform be $m\inr$, $g$ be the gravitational constant, $\bs{e}_z = [0 \ 0 \ 1]^\top$, and $\bs{\sigma}_\rho \inr^3$ be the lumped translational disturbance term which satisfies \textit{Assumption \ref{as:disturbance}}. Then, the translational dynamics~\cite{lee2010geometric} are
\begin{equation}\label{eq:translationalDynamics}
    \begin{split}
        \dot{\bs{p}}_1 &= \bs{p}_2 \\
        \dot{\bs{p}}_2 &= -\frac{v_f}{m}R_3(\bs{\vartheta}_1) + g\bs{e}_z + \bs{\sigma}_\rho,
    \end{split}
\end{equation}
where
\begin{equation}
    R_3(\bs{\vartheta}_1) = \left[\begin{matrix}
    \cos(\phi) \sin(\theta) \cos(\psi) + \sin(\phi) \sin(\psi)\\
    \cos(\phi) \sin(\theta) \sin(\psi) - \sin(\phi) \cos(\psi)\\
    \cos(\phi) \cos(\theta) \end{matrix}\right].
\end{equation}

Let $\bs{p}_r = [x_r \ y_r \ z_r]^\top \inr^3$ and $\dot{\bs{p}}_r = [\dot{x}_r \ \dot{y}_r \ \dot{z}_r]^\top \inr^3$ be the translational reference signals. Define the translational error variables
\begin{equation}
    \begin{gathered}
        \bs{\rho}_1 = \bs{p}_1 - \bs{p}_r, \quad \bs{\rho}_2 = \dot{\bs{\rho}_1} = \bs{p}_2 - \dot{\bs{p}}_r, \quad \bs \rho=[\bs \rho_1^\top \; \bs \rho_2^\top]^\top.
    \end{gathered}
\end{equation}
The translational dynamics \eqref{eq:translationalDynamics} can now be written in terms of tracking error as
\begin{equation}\label{eq:translationalErrorDynamics}
    \begin{split}
        \dot{\bs{\rho}}_1 &= \bs{\rho}_2 \\
        \dot{\bs{\rho}}_2 &= -\frac{v_f}{m}R_3(\bs{\vartheta}_1) + g\bs{e}_z + \bs{\sigma}_\rho - \ddot{\bs{p}}_r.
    \end{split}
\end{equation}

\subsection{Reference System Dynamics}
We assume that the reference trajectory that the multi-rotor UAV will track is generated by the system
\begin{equation}\label{eq:groundVehicleDynamics}
    \begin{split}
        \dot{\bs{p}}_{c_1} &= \bs{p}_{c_2} \\
        \dot{\bs{p}}_{c_2} &= f_c(\bs{p}_c,\bs{u}_c), \\
    \end{split}
\end{equation}
where $\bs{p}_{c_1} = [x_c \ y_c \ z_c]^\top \inr^3$ and $\dot{\bs{p}}_{c_1} = [\dot{x}_c \ \dot{y}_c \ \dot{z}_c]^\top \inr^3$ are the position and velocity of the reference system, $\bs p_c =[\bs p_{c_1}^\top, \bs p_{c_2}^\top]^\top$ is the system state, $\bs u_c$ is the unknown system input, and $f_c(\bs{p}_c,\bs{u}_c)$ is some unknown function. We take the system input $\bs{u}_c = g_c(t,\bs{p}_c)$ and let $\bar{f}_c(t,\bs{p}_c) = f_c(\bs{p}_c,\bs{u}_c)$.  We assume that $\frac{\partial \bar{f}_c(t,\bs{p}_c)}{\partial \bs{p}_c}\dot{\bs{p}}_c$ satisfies \textit{Assumption \ref{as:disturbance}}. In the case of tracking a moving ground vehicle, the reference signals will be taken as the reference system state, $\bs{p}_c$, and will be estimated using measurements of the ground vehicle position. We initially describe a generic reference signal, $\bs{p}_r$, to keep the control design general. We then take these dynamics as our reference signal in the specific case of landing on a ground vehicle.

\subsection{Actuator Dynamics and Mapping to Inputs}
The system dynamics, \eqref{eq:rotationalErrorDynamics} and \eqref{eq:translationalErrorDynamics}, take body-fixed torques, $\bs{\tau}$, and total thrust force, $v_f$, as inputs. The thrust and torques are generated by applying forces with each actuator. The force generated by rotor $i \in \until{n}$ is $\bar{f}_i = C_T\omega_i^2$, where $C_T \inr_{>0}$ is a constant relating angular rate to force and $\omega_i \inr_{>0}$ is the $i$-th rotor angular rate. These individual actuator forces are then mapped through a matrix, $M \inr^{4\times n}$, based on the geometry of the multi-rotor aerial platform, allowing the squared rotor angular rates to be treated as the system input through
\begin{equation}\label{eq:inputTransformation}
    \left[\begin{matrix} v_f \\ \bs{\tau} \end{matrix}\right]= C_T M\bs{\omega}_s, \quad \bs{\omega}_s = \left[\omega_1^2, \dots, \omega_n^2 \right]^\top.
\end{equation}

The actuators typically used on multi-rotor UAVs are Brushless DC (BLDC) motors, which require electronic speed controllers (ESCs). Let the vector of desired rotor angular rates be $\bs{\omega}_\text{des} \inr^n$ and $\bs{\omega} \inr^n$ be the vector of rotor angular rates. Due to the internal use of PI control in the ESCs \cite{franchi2017adaptive}, the rotor angular rates exhibit first-order dynamics of the following form
\begin{equation}\label{eq:actuatorDynamics}
    \tau_m \dot{\bs{\omega}} = (\bs{\omega}_\text{des} - \bs{\omega}),
\end{equation}
where $\tau_m \inr_{>0}$ is the time constant of the actuator system. Typically the actuator dynamics are ignored in multi-rotor control design as they are sufficiently fast compared with the rotational and translational dynamics and the control law. We also ignore the actuator dynamics in our control design, however, they are crucial in the dynamics of the EHGO used for output feedback control (see Section \ref{rem:actuatorDynamics} below). The actuator dynamics evolve on the same timescale as the EHGO dynamics, and therefore cannot be ignored in EHGO design.

Since measurement of the rotor angular rates is not available, they can be simulated by the following system
\begin{equation}\label{eq:simulatedActuatorDynamics}
    \tau_m\dot{\hat{\bs{\omega}}} = (\bs{\omega}_\text{des} - \hat{\bs{\omega}}), \quad \hat{\omega}(0) = 0_{n\times1},
\end{equation}
where $\hat{\bs{\omega}}\inr^n$ is a vector of simulated rotor angular rates, and $0_{n\times1}\inr^{n\times1}$ is a vector of zeros. We will show in Section \ref{sec:stabilityAnalysis} that the use of simulated rotor speeds in place of measured rotor speeds still results in an exponentially stable closed-loop system.

\section{Control and Observer Design} \label{sec:ControlDesign}
A multi-rotor UAV is an underactuated mechanical system. While there can be $n\in\{4,6,8,...\}$ rotors, only four degrees of freedom can be controlled in the classic configuration with co-planar rotors. To overcome the under-actuation, as discussed below,  the rotational dynamics are controlled to create a virtual control input for the translational dynamics.

We begin by designing a trajectory-tracking feedback linearizing controller for the rotational subsystem. The rotational trajectory is subsequently used to design a trajectory-tracking controller for the translational subsystem in the presence of tracking errors in the rotational system. The controllers are designed under state feedback which requires the assumption that we not only have access to all states but know the system disturbances exactly. This assumption is relaxed through the design of an EHGO to estimate states, disturbances, and the reference trajectory for use in output feedback control.

\subsection{Rotational Control}
The rotational control feedback linearizes the rotational tracking error dynamics \eqref{eq:rotationalErrorDynamics} by selecting the desired torque $\bs{\tau}$ as
\begin{equation}\label{eq:rotationalControlSF}
        \bs{\tau}_d = G^{-1}(\bs{\vartheta}_1)[\bs{u}_r - f(\bs{\xi},\bs{\vartheta}_1,\dot{\bar{\bs{\vartheta}}}_r)],
\end{equation}
where $\bs{u}_r = -\beta_1\bs{\xi}_1 - \beta_2\bs{\xi}_2 - \bs{\varsigma}_\xi$, and $\beta_1,\beta_2\inr_{>0}$ are constant gains. Using \eqref{eq:rotationalControlSF} results in the following closed-loop rotational tracking error system
\begin{equation}\label{eq:rotationalClosedLoopStateFeedback}
    \begin{split}
        \dot{\bs{\xi}}_1 &= \bs{\xi}_2\\
        \dot{\bs{\xi}}_2 &= -\beta_1\bs{\xi}_1 - \beta_2\bs{\xi}_2.
    \end{split}
\end{equation}

\subsection{Translational Control}
The translational control uses the total thrust, $v_f$, as the direct control input and the desired roll and pitch trajectories, $\phi_r$ and $\theta_r$, as virtual control inputs. The translational control is designed in view of potential roll and pitch trajectory tracking errors, leading to the following modification of the translational error dynamics \eqref{eq:translationalErrorDynamics}
\begin{equation}\label{eq:translationalRotationalErrorDynamics}
    \begin{split}
        \dot{\bs{\rho}}_1 &= \bs{\rho}_2 \\
        \dot{\bs{\rho}}_2 &= -\frac{v_f}{m}R_3(\bs{\vartheta}_r + \bs{\xi}_1) + g\bs{e}_z + \bs{\sigma}_\rho - \ddot{\bs{p}}_r.
    \end{split}
\end{equation}
Define the perturbation due to rotational tracking error by
\begin{equation}\label{eq:translationalRotationalErrorTerm}
    \bs{e}_\vartheta(t,\bs{\xi}_1) = -\frac{v_f}{m}(R_3(\bs{\vartheta}_r + \bs{\xi}_1) - R_3(\bs{\vartheta}_r)).
\end{equation}
While we establish that there is no longer a time-scale separation required between the rotational and translational subsystems, in practice the rotational control should be at least slightly faster than the translational dynamics to provide good tracking performance when operating in the relatively slow sample rates achievable by our off-the-shelf hardware. We will establish that all control inputs remain bounded provided the initial conditions are in an appropriate set which will be characterized in Section \ref{sec:stabilityAnalysis}, therefore \eqref{eq:translationalRotationalErrorTerm} will remain bounded as well.
Then, \eqref{eq:translationalRotationalErrorDynamics} can be written as
\begin{equation}\label{eq:translationalDynamicsWithErrorTerm}
    \begin{split}
        \dot{\bs{\rho}}_1 &= \bs{\rho}_2 \\
        \dot{\bs{\rho}}_2 &= -\frac{v_f}{m}R_3(\bs{\vartheta}_r) + g\bs{e}_z + \bs{\sigma}_\rho - \ddot{\bs{p}}_r + \bs{e}_\vartheta(t,\bs{\xi}_1).
    \end{split}
\end{equation}

Let $\bs{u}_t = [u_x \ u_y \ u_z]^\top$ be defined by $\bs{u}_t = -\gamma_1\bs{\rho}_1 - \gamma_2\bs{\rho}_2 - \bs{\sigma}_\rho + \ddot{\bs{p}}_r - g\bs{e}_z$, where $\gamma_1,\gamma_2 \inr_{>0}$ are constant gains. Define the desired rotational references and desired total thrust by
\begin{equation}\label{eq:translationalControlSF}
    \begin{split}
        \phi_r &= \tan^{-1}\left( \dfrac{-u_y}{\sqrt{u_x^2 + u_z^2}} \right), \quad \psi_r = 0,\\ 
        \theta_r &= \tan^{-1}\left(\dfrac{u_x}{u_z}\right), \quad
        v_{f d} = -\dfrac{mu_z}{\cos(\phi_r) \cos(\theta_r)}.
    \end{split}
\end{equation}
Then, $-\frac{v_f}{m} R_3(\bs{\vartheta}_r) = \bs{u}_t$. Thus, using \eqref{eq:translationalControlSF} leads to the following closed-loop translational subsystem with the inclusion of tracking error \eqref{eq:translationalRotationalErrorTerm} from the rotational subsystem
\begin{equation}\label{eq:translationalClosedLoopStateFeedback}
    \begin{split}
        \dot{\bs{\rho}}_1 &= \bs{\rho}_2 \\
        \dot{\bs{\rho}}_2 &= -\gamma_1\bs{\rho}_1 - \gamma_2\bs{\rho}_2 + \bs{e}_\vartheta(t,\bs{\xi}_1).
    \end{split}
\end{equation}
Note that the rotational controller \eqref{eq:rotationalControlSF} requires the estimate $\dot{\bar{\bs{\vartheta}}}_r$, however, only $\bs{\vartheta}_r$ is given by the translational controller \eqref{eq:translationalControlSF}. The derivative of the reference trajectory $\dot{\bs{\vartheta}}_r$ can be computed analytically from the translational controller as
\begin{equation}\label{eq:thetaRdot}
    \begin{split}
        \dot{\phi}_r &= \frac{ u_y\left(\dot{u}_x u_x + \dot{u}_z u_z\right) - \dot{u}_y\left(u_x^2 + u_z^2\right)}{\left(u_x^2+u_z^2\right)^{1/2}\left(u_x^2 + u_y^2 + u_z^2\right)}\\
        \dot{\theta}_r &= \frac{\dot{u}_x u_z - u_x \dot{u}_z}{u_x^2 + u_z^2} \\
        \dot{\psi}_r &= 0,
    \end{split}
\end{equation}
where $\dot{\bs{u}}_t = \left[\dot{u}_x \ \dot{u}_y \ \dot{u}_z\right]^\top$ and
\begin{equation}\label{eq:u_tDot}
    \begin{split}
    \dot{\bs{u}}_t &= -\gamma_1\bs{\rho}_2 - \gamma_2\left[-\frac{v_f}{m}R_3(\bs{\vartheta}_1) + g\bs{e}_z + \bs{\sigma}_\rho - \ddot{\bs{p}}_r \right] \\
    &\quad - \dot{\bs{\sigma}}_\rho + \bs{p}_r^{(3)}.
    \end{split}
\end{equation}

The approximation $\dot{\bar{\bs{\vartheta}}}_r$ is obtained by setting $\dot{\bs{\sigma}}_\rho = 0$ in the expression for $\dot{\bs{\vartheta}}_r$. While the derivative of the translational disturbance is most certainly not zero, we try to capture the behavior of $\dot{\bs{\vartheta}}_r$ as closely as we can. We have no information on how $\dot{\bs{\sigma}}_\rho$ behaves, so we choose to set it to zero in the approximation. The implication is an increase in the magnitude of the rotational disturbance term $\bs{\varsigma}_\xi$, see Remark \ref{rem:referenceApproximation}. Recall that our definition of state feedback assumes $\bs{\sigma}_\rho$ is known, and later in the high-gain observer we will estimate $\bs{\sigma}_\rho$ and we will analyze the effect of uncertainty in the estimate on feedback performance. While the substitution \eqref{eq:u_tDot} requires the third-order derivative of the translational reference, it is shown in the EHGO design that the translational reference must be sixth-order differentiable to be sufficiently smooth for estimation. 

\subsection{Extended High-Gain Observer Design}\label{sec:EHGODesign}
A multi-input multi-output EHGO is designed similar to \cite{lee2012control,lee2016output} to estimate higher-order states of the error dynamic systems \eqref{eq:rotationalErrorDynamics} and \eqref{eq:translationalErrorDynamics}, uncertainties arising from modeling error and external disturbances, as well as the reference trajectory based on the reference system dynamics \eqref{eq:groundVehicleDynamics}. It is shown in \cite{boss2017uncertainty} that the actuator dynamics must be included in the dynamic model in the EHGO design.

The dynamics \eqref{eq:rotationalDynamics}, \eqref{eq:translationalDynamics}, and \eqref{eq:groundVehicleDynamics} can be combined into one set of equations for the observer where the state space is extended to include unknown disturbance dynamics. Since the third derivative of the reference trajectory is required by \eqref{eq:u_tDot}, the dynamics of the reference system are extended to include the third derivative of its position for estimation
\begin{equation}
    \begin{split}
    \dot{\bs{\rho}}_1 &= \bs{\rho}_2 \\
        \dot{\bs{\rho}}_2 &= -\frac{v_f}{m}R_3(\bs{\vartheta}_1) + g\bs{e}_z + \bs{\sigma}_\rho - \ddot{\bs{p}}_r \\
        \dot{\bs{\sigma}}_\rho &= \varphi_\rho(t,\bs{\rho})\\
        \dot{\bs{\xi}}_1 &= \bs{\xi}_2 \\
        \dot{\bs{\xi}}_2 &= f(\bs{\xi},\bs{\vartheta}_1,\dot{\bar{\bs{\vartheta}}}_r) + G(\bs{\vartheta}_1)\bs{\tau} + \bs{\varsigma}_\xi \\
        \dot{\bs{\varsigma}}_\xi &= \varphi_\xi(t,\bs{\xi})\\
        \dot{\bs{p}}_{c_1} &= \bs{p}_{c_2} \\
        \dot{\bs{p}}_{c_2} &= \bs{p}_{c_3} \\
        \dot{\bs{p}}_{c_3} &= \bs{\sigma}_{p c} \\
        \dot{\bs{\sigma}}_{p c} &= \varphi_{p c}(t,\bs{p}_c),\\
    \end{split}
\end{equation}
where $ \bs{\sigma}_{pc} =\frac{\partial \bar{f}_c(t,\bs{p}_c)}{\partial \bs{p}_c}\dot{\bs{p}}_c$. Since the reference system dynamics may not be known, they have been absorbed by the disturbance term in their entirety. If the reference system dynamics are partially known, then the nominal component can be included in the $\dot{\bs{p}}_{c_3}$ expression. The estimated reference system states will be taken as the reference trajectory for the output feedback control.

We now define the state vectors 
\begin{equation}
    \begin{gathered}
        \bs{q} = [\bs{\rho}_1^\top \; \bs{\rho}_2^\top \; \bs{\xi}_1^\top \; \bs{\xi}_2^\top]^\top, \quad \bs{\chi}_1 = [\bs{\rho}_1^\top \; \bs{\rho}_2^\top \; \bs{\sigma}_\rho^\top]^\top, \\
        \bs{\chi}_2 = [\bs{\xi}_1^\top \; \bs{\xi}_2^\top \; \bs{\varsigma}_\xi^\top]^\top, \quad \bs{\chi}_3 = [\bs{p}_{c_1}^\top \; \bs{p}_{c_2}^\top \; \bs{p}_{c_3}^\top \;
        \bs{\sigma}_{p c}^\top]^\top,\\
\text{and} \quad        \bs{\chi} = [\bs{\chi}_1^\top \; \bs{\chi}_2^\top \; \bs{\chi}_3^\top]^\top.
    \end{gathered}
\end{equation}
Define $\varphi(t,\bs{q},\bs{p}_c) = \left[\varphi_\rho(t,\bs{\rho}) \ \varphi_\xi(t,\bs{\xi}) \ \varphi_{p c}(t,\bs{p}_c)\right]^\top$, a vector of unknown functions describing the disturbance dynamics. 
\begin{assumption}[Disturbance Dynamics]\label{as:disturbanceDynamics}
    It is assumed $\varphi(t,\bs{q},\bs{p}_c)$ is continuous and bounded on any compact set containing $\bs{q}$ and $\bs{p}_c$.
\end{assumption} 

Note that the second order derivative of the reference trajectory, $\ddot{\bs{\vartheta}}_r$, is lumped into the disturbance $\bs{\varsigma}_\xi$. To ensure $\bs{\varsigma}_\xi$ satisfies \textit{Assumption \ref{as:disturbance}}, $\ddot{\bs{\vartheta}}_r$ must be differentiable, therefore by \eqref{eq:thetaRdot} and \eqref{eq:u_tDot} the translational reference signals must be sixth order differentiable to be sufficiently smooth. We are also assuming the external disturbance $\bs{\sigma}_\rho$ is sufficiently smooth since it is contained within $\ddot{\bs{\vartheta}}_r$. However, our design only requires estimates up to the third derivative. 

The observer system with extended states and a vector of simulated squared rotor speeds, $\hat{\bs{\omega}}_s = [\hat{\omega}_1^2, \dots, \hat{\omega}_n^2]^\top$ from the system \eqref{eq:simulatedActuatorDynamics}, as the control input through the mapping \eqref{eq:inputTransformation} is
\begin{equation}\label{eq:fullObserverDynamics}
    \begin{split}
        \dot{\hat{\bs{\chi}}} &= A\hat{\bs{\chi}} + B\left[\bar{f}(\hat{\bs{\xi}},\hat{\bs{p}}_{c_3},\bs{\vartheta}_1,\dot{\bar{\bs{\vartheta}}}_r) + \bar{G}(\bs{\vartheta}_1)\hat{\bs{\omega}}_s\right] + H\hat{\bs{\chi}}_e \\
        \hat{\bs{\chi}}_e &= C(\bs{\chi} - \hat{\bs{\chi}}),
    \end{split}
\end{equation}
where

\begin{subequations}
\begin{equation}
    \begin{gathered}
        A = \oplus_{i=1}^3 A_i, \ B = \oplus_{i=1}^3 B_i, \ 
        C = \oplus_{i=1}^3 C_i, \ H = \oplus_{i=1}^3 H_i,
    \end{gathered}
\end{equation}
\begin{equation}
    \begin{gathered}
        A_i = \left[\begin{smallmatrix} 0_3 & I_3 & 0_3 \\ 0_3 & 0_3 & I_3 \\ 0_3 & 0_3 & 0_3\end{smallmatrix}\right], \ B_i = \left[\begin{smallmatrix} 0_3\\ I_3 \\ 0_3\end{smallmatrix}\right], \ H_i = \left[\begin{smallmatrix} \alpha_1/\epsilon I_3 \\ \alpha_2/\epsilon^2 I_3 \\ \alpha_3/\epsilon^3 I_3 \end{smallmatrix}\right], \\
        C_i = \left[\begin{smallmatrix} I_3 & 0_3 & 0_3\end{smallmatrix}\right], \ \text{for} \ i \in \{1,2\}, 
    \end{gathered}
\end{equation}
\begin{equation}
    \begin{gathered}
        A_3 = \left[\begin{smallmatrix} 0_3 & I_3 & 0_3 & 0_3\\ 0_3 & 0_3 & I_3 & 0_3\\ 0_3 & 0_3 & 0_3 & I_3 \\ 0_3 & 0_3 & 0_3 & 0_3\end{smallmatrix}\right], \ B_3 = \left[\begin{smallmatrix} 0_3 \\ 0_3 \\ 0_3 \\ 0_3\end{smallmatrix}\right], \ H_3 = \left[\begin{smallmatrix} \alpha_1/\epsilon I_3 \\ \alpha_2/\epsilon^2 I_3 \\ \alpha_3/\epsilon^3 I_3 \\ \alpha_4/\epsilon^4 I_3\end{smallmatrix}\right], \\
        C_3 = \left[\begin{smallmatrix} I_3 & 0_3 & 0_3 & 0_3\end{smallmatrix}\right],
    \end{gathered}
\end{equation}
\begin{equation}
    \bar{f}(\hat{\bs{\xi}},\hat{\bs{p}}_{c_3},\bs{\vartheta}_1,\dot{\bar{\bs{\vartheta}}}_r) = \left[\begin{smallmatrix} g\bs{e}_z - \hat{\bs{p}}_{c_3} \\ f(\hat{\bs{\xi}},\bs{\vartheta}_1,\dot{\bar{\bs{\vartheta}}}_r) \\ 0_{3\times1}\end{smallmatrix}\right],
\end{equation}
\begin{equation}
    \bar{G}(\bs{\vartheta}_1) = C_T\left[\begin{smallmatrix} \frac{-R_3(\bs{\vartheta}_1)}{m} & 0_{3} \\ 0_{3\times1} & G(\bs{\vartheta}_1) \\ 0_{3\times1} & 0_3 \end{smallmatrix}\right]M,
\end{equation}
\end{subequations}
where $\oplus$ denotes the matrix direct sum, $I_n\inr^{n\times n}$ is the identity matrix of dimension $n$, $0_n\inr^{n\times n}$ is a square matrix of zeros, and $H$ is designed by choosing $\alpha_j^i$ such that
\begin{equation}
    s^{\varrho_i} + \alpha_1^i s^{\varrho_i-1} + \dots + \alpha_{\varrho_i-1}^i s + \alpha_{\varrho_i},
\end{equation}
is Hurwitz, in this case $[\varrho_1 \ \varrho_2 \ \varrho_3]^\top = [3 \ 3 \ 4]^\top$, which are chosen through tuning on the experimental platform, and $\epsilon\inr_{>0}$ is a positive constant that is chosen small enough. Note, based on the structure of $C$, $\chi_e$ only contains $[\rho_1^T , \xi_1^T , p_{c1}^T ]^T$.

\subsection{Output Feedback Control}
For use in output feedback control, the estimates, $\hat{\bs{\chi}}$, must be saturated outside a compact set of interest to overcome the peaking phenomenon (see \textit{Appendix \ref{A:peakingPhenomenon}}). The following saturation function is used to saturate each estimate individually
\begin{equation}\label{eq:estimateSaturation}
    \hat{\chi}_{is} = k_{\chi_i}\operatorname{sat}\left(\frac{\hat{\chi}_i}{k_{\chi_i}} \right), \ \  \operatorname{sat}(y)=\left\{\begin{array}{ll}{y,} & {\text { if }|y| \leq 1}, \\ {\operatorname{sign}(y),} & {\text { if }|y|>1},\end{array}\right.
\end{equation}
for $1 \leq i \leq 30$, where the saturation bounds $k_{\chi_i}$ are chosen such that the saturation functions will not be invoked under state feedback.

The state feedback controllers \eqref{eq:rotationalControlSF} and \eqref{eq:translationalControlSF} are rewritten as output feedback controllers using the saturated estimates
\begin{equation}\label{eq:rotationalOutputFeedback}
    \hat{\bs{\tau}}_d = G^{-1}(\bs{\vartheta}_1)\left[\hat{\bs{u}}_r - f(\hat{\bs{\xi}},\bs{\vartheta}_1,\dot{\bar{\bs{\vartheta}}}_r)\right],
\end{equation}
where $\hat{\bs{u}}_r = -\beta_1\hat{\bs{\xi}}_1 - \beta_2\hat{\bs{\xi}}_2 - \hat{\bs{\varsigma}}_\xi$ and
\begin{equation}\label{eq:translationalOutputFeedback}
    \begin{split}
        \hat{\phi}_r &= \tan^{-1}\left( \dfrac{-\hat{u}_y}{\sqrt{\hat{u}_x^2 + \hat{u}_z^2}} \right), \quad \hat{\psi}_r = 0, \\
        \hat{\theta}_r &= \tan^{-1}\left(\dfrac{\hat{u}_x}{\hat{u}_z}\right), \quad \hat{v}_{f d} = -\dfrac{m\hat{u}_z}{\cos(\hat{\phi}_r) \cos(\hat{\theta}_r)},
    \end{split}
\end{equation}
where $\hat{\bs{u}}_t = -\gamma_1\hat{\bs{\rho}}_1 - \gamma_2\hat{\bs{\rho}}_2 - \hat{\bs{\sigma}}_\rho + \hat{\bs{p}}_{c_3} - g\bs{e}_z$.

Furthermore, these control inputs can be mapped to desired squared rotor speeds, $\bs{\omega}_{s d} \inr^n$, from the output feedback linearizing control signals $\hat{v}_{f d}$ and $\hat{\bs{\tau}}_d$. For $n>4$, the inverse of \eqref{eq:inputTransformation} is an over-determined system that admits infinitely many solutions. In this case, we focus on the minimum energy solution
\begin{equation}\label{eq:omegaSD}
    \bs{\omega}_{s d} = \frac{1}{C_T}M^\dagger \left[\begin{matrix} \hat{v}_{f d} \\ \hat{\bs{\tau}}_d \end{matrix}\right], \quad \text{where} \; M^\dagger = M^\top (MM^\top)^{-1}.
\end{equation}
The square root of each component of $\bs{\omega}_{s d}$ acts as the reference signal, $\bs{\omega}_\text{des}$, in \eqref{eq:simulatedActuatorDynamics} for the associated rotor, which in turn can be applied directly to the physical system.

The overall output feedback controller consists of the commanded three body-fixed torques, $\hat{\bs{\tau}}_d$, and the commanded collective thrust, $\hat{v}_{f d}$, restated here for ease of reference.
\begin{equation}
    \begin{split}
        \hat{\bs{\tau}}_d &= G^{-1}(\bs{\vartheta}_1)\left[\hat{\bs{u}}_r - f(\hat{\bs{\xi}},\bs{\vartheta}_1,\dot{\bar{\bs{\vartheta}}}_r)\right], \\
        \hat{v}_{f d} &= -\dfrac{m\hat{u}_z}{\cos(\hat{\phi}_r) \cos(\hat{\theta}_r)}.
    \end{split}
\end{equation}

\subsection{Illustration of Instability Induced by Ignoring Actuator Dynamics}\label{rem:actuatorDynamics}
When designing an EHGO for highly dynamic systems, the inclusion of faster system dynamics must be considered. In the case of multi-rotors, the actuator dynamics evolve on the same timescale as the EHGO. If these dynamics are not considered in the EHGO design, overall system stability can be compromised. Consider an EHGO that neglects the actuator dynamics. While the actuators are changing their rotational rates according to \eqref{eq:actuatorDynamics} to apply the desired control input, the EHGO, with no knowledge of these relatively slow dynamics, will observe this delayed application of control as a large disturbance. In an effort to cancel this perceived disturbance, a larger control action is commanded. This causes the system to overshoot the reference dramatically. The opposite action occurs in trying to correct the overshoot, resulting in aggressive oscillations that can destabilize the system.

\begin{figure}
    \centering
    \includegraphics[width=0.47\textwidth]{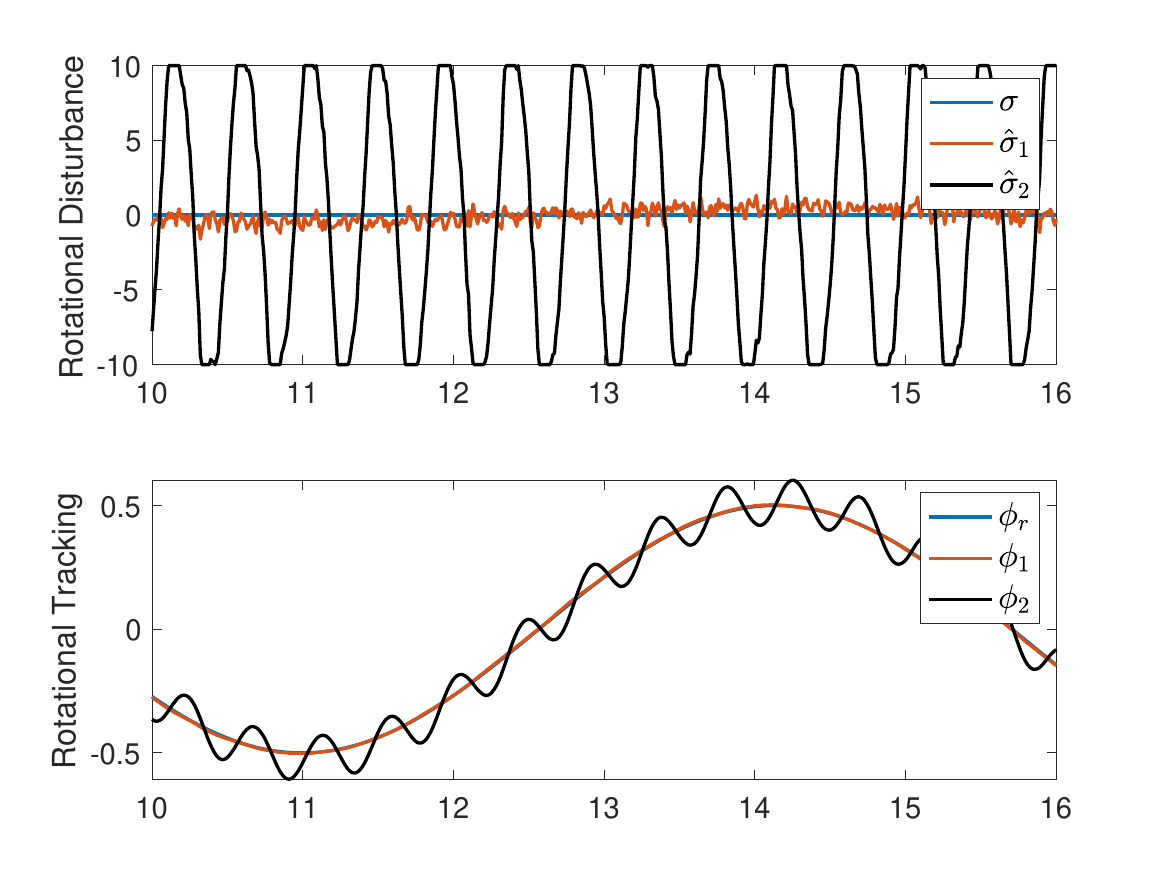}
    \caption{Simulated rotational system response with and without actuator dynamics included in the EHGO. The disturbance estimate, $\hat{\sigma}_2$, when actuator dynamics are omitted oscillates between the saturation bounds (top), inducing oscillations in the tracking performance of $\phi_2$ (bottom). The disturbance estimate, $\hat{\sigma}_1$, and tracking, $\phi_1$, show excellent performance when actuator dynamics are included in the EHGO for this example.}
    \label{fig:actuatorDyanmicEffect}
\end{figure}

This behavior is illustrated in Fig. \ref{fig:actuatorDyanmicEffect}, where the rotational subsystem is simulated with and without actuator dynamics in the observer. There is no nominal disturbance applied to the system, however, the disturbance estimate from the observer without actuator dynamics oscillates quickly between its saturation bounds. In this case, the saturation bounds were chosen small enough to prevent the system from becoming unstable to illustrate the oscillatory behavior induced by the omission of the actuator dynamics. When the EHGO has a model of how the actuators are dynamically applying the desired control action, there is no longer a perceived disturbance due to the actuator dynamics, and the system functions nominally. Through simulation, it was found that observer performance is only mildly sensitive to the time constant of the actuator dynamics. Performance is recovered so long as the time constant of the dynamic model included in the observer is within a factor of two of the true dynamics.

Indeed, in the presence of disturbance input, the EHGO can estimate the system state with $O(\epsilon \varDelta_{\max})$\footnote{Here, $O(\star)$ means the signal is of order $\star$. In other words, the signal is upper bounded by some constant multiplied by $\star$.} accuracy, where $\varDelta_{\max}$ is an upper bound on the size of the disturbance~\cite{khalil2017HGO}. The maximum available measurement sampling rate imposes a lower bound on $\epsilon$. Even a partially accurate model of actuator dynamics reduces $\varDelta_{\max}$ and leads to superior EHGO performance. Therefore, including the actuator dynamics is helpful even if the associated parameters are not perfectly known.

\section{Stability Analysis}\label{sec:stabilityAnalysis}
In this section, we will derive the requirements of the initial conditions that ensure that the proposed controller is well-defined throughout the operation. We then establish the stability of the state feedback control, observer estimates, and output feedback control.
\subsection{Restricting Domain of Operation}
The domain of operation must be restricted in order to ensure that the rotational feedback linearizing control law remains well-defined. To ensure the expressions in \eqref{eq:translationalControlSF} are well-defined, we introduce the following assumption.
\begin{assumption}\label{as:rotationalRefSet}
The rotational reference signals remain in the set $\{|\phi_r| < \frac{\pi}{2} - \delta, |\theta_r| < \frac{\pi}{2} - \delta, |\psi_r| < \frac{\pi}{2} - \delta \}$, where $0<\delta<\frac{\pi}{2}$.
\end{assumption}

To ensure the rotational tracking error is well defined, i.e., the magnitude of each entry of $\bs{\xi}_1$ is smaller than $\frac{\pi}{2}$, and to ensure singularities of the Z-Y-X Euler angle representation at $\theta = \pm \frac{\pi}{2}$, the rotational states must remain in the set $\{|\phi| < \frac{\pi}{2}, |\theta| < \frac{\pi}{2}, |\psi| < \frac{\pi}{2}, |\dot{\phi}| < a_\theta, |\dot{\theta}| < a_\theta, |\dot{\psi}| < a_\theta \}$, where $a_\theta$ is some positive constant. The magnitude of each entry of $\bs{\xi}_1$ should be smaller than $\frac{\pi}{2}$ to ensure that the rotational error is well-defined. We will now establish that for sufficiently small initial tracking error, $\bs{\xi}(0)$, the tracking error $\norm{\bs{\xi}_1(t)}<\delta$ for all $t>0$. A Lyapunov function in rotational error dynamics is
\begin{subequations}\label{eq:v_xi}
\begin{equation}\label{eq:v_xi_bnd}
    V_\xi = \bs{\xi}^\top P_\xi\bs{\xi}, \quad \text{where} \ P_\xi A_\xi + A_\xi^\top P_\xi = -I_6,
\end{equation}
\begin{equation}
    A_\xi = \left[\begin{smallmatrix} 0_3 & I_3 \\ -\beta_1I_3 & -\beta_2I_3 \end{smallmatrix}\right].
\end{equation}
\end{subequations}
A Lyapunov function in the translational error dynamics is
\begin{subequations}
\begin{equation}\label{eq:v_rho}
    V_\rho = \bs{\rho}^\top P_\rho\bs{\rho}, \quad \text{where} \ P_\rho A_\rho + A_\rho^\top P_\rho = -I_6,
\end{equation}
\begin{equation}
    A_\rho = \left[\begin{smallmatrix} 0_3 & I_3 \\ -\gamma_1I_3 & -\gamma_2I_3 \end{smallmatrix}\right].
\end{equation}
\end{subequations}
Solving $P_\xi A_\xi + A_\xi^\top P_\xi = -I_6$ for $P_\xi$ and $P_\rho A_\rho + A_\rho^\top P_\rho = -I_6$ for $P_\rho$ yields
\begin{equation}
    P_\xi = \left[\begin{smallmatrix} \frac{\beta_1^2 + \beta_1 + \beta_2^2}{2\beta_1\beta_2}I_3 & \frac{1}{2\beta_1}I_3 \\ \frac{1}{2\beta_1}I_3 & \frac{\beta_1 + 1}{2\beta_1\beta_2}I_3 \end{smallmatrix}\right], \quad P_\rho = \left[\begin{smallmatrix} \frac{\gamma_1^2 + \gamma_1 + \gamma_2^2}{2\gamma_1\gamma_2}I_3 & \frac{1}{2\gamma_1}I_3 \\ \frac{1}{2\gamma_1}I_3 & \frac{\gamma_1 + 1}{2\gamma_1\gamma_2}I_3 \end{smallmatrix}\right].
\end{equation}
Let $c_\xi\inr_{>0}$ be chosen such that $c_\xi < (\beta_1 + 1)\delta^2/(2\beta_2)$, and let $\Omega_\xi = \{V_\xi < c_\xi\}$. Since $\bs{e}_\vartheta(t,\bs{\xi}_1)$ and its partial derivatives are continuous on $\Omega_\xi$, and $\bs{e}_\vartheta$ is uniformly bounded in time, it is locally Lipschitz in $\Omega_\xi$ and let $L_e$ be the associated Lipschitz constant. Take $c_\rho > \lambda_\text{max}(P_\rho)(2L_e\delta\lambda_\text{max}(P_\rho))^2$, where $\lambda_\text{max}(\cdot)$ is the maximum eigenvalue of the argument,
and let $\Omega_\rho = \{V_\rho < c_\rho\}$. Define the domain of operation $\Omega_q = \Omega_\xi \times \Omega_\rho$.
\begin{lemma}[Restricting Domain of Operation]\label{lem:restrictingDomainOfAnalysis}
    For the feedback linearized rotational error dynamics \eqref{eq:rotationalClosedLoopStateFeedback} with initial conditions $\bs{\xi}(0) \in \Omega_\xi$ the system state $\bs{\xi}(t)$ remains in the set $\norm{\bs{\xi}_1(t)}<\delta$ for all $t>0$. Similarly, the feedback linearized translational error dynamics \eqref{eq:translationalClosedLoopStateFeedback} with initial conditions $\bs{\rho}(0) \in \Omega_\rho$, the system state $\bs{\rho}(t)$ remains in $\Omega_\rho$, for all $t>0$.
\end{lemma}
\begin{proof}
    See Appendix \ref{A:proofRestrictingDomain}.
\end{proof}

\begin{remark}\label{rem:rotationalSet}
By \textit{Lemma \ref{lem:restrictingDomainOfAnalysis}} and \textit{Assumption \ref{as:rotationalRefSet}}, the rotational states remain in the set $\{|\phi| < \frac{\pi}{2}, |\theta| < \frac{\pi}{2}, |\dot{\phi}| < a_\theta, |\dot{\theta}| < a_\theta \}$, where $a_\theta$ is some positive constant. Thereby ensuring singularities in the Euler angles are avoided and the feedback linearizing controllers \eqref{eq:rotationalControlSF} and \eqref{eq:translationalControlSF} remain well defined.
\end{remark}

Furthermore, we will restrict the domain of operation of the reference system by defining the set $\Omega_{p_c} = \{\|\bs{p}_{c_1}\| < a_1, \|\bs{p}_{c_2}\| < a_2, \|\bs{p}_{c_3}\| < a_3\}$ for $a_1,a_2,a_3\inr_{>0}$.

\subsection{Stability Under State Feedback}\label{subsec:stabilityUnderStateFeedback}
\begin{theorem}[Stability Under State Feedback]\label{thm:SFStability}
    For the closed-loop state feedback rotational and translational subsystems,  \eqref{eq:rotationalClosedLoopStateFeedback} and \eqref{eq:translationalClosedLoopStateFeedback}, if the initial conditions $(\bs{\xi}(0),\bs{\rho}(0)) \in \Omega_q$, the system states $(\bs{\xi}(t),\bs{\rho}(t)) \in \Omega_q$ for all $t>0$. Additionally, the states will exponentially converge to the origin.
\end{theorem}
\begin{proof}
    See Appendix \ref{A:proofStateFeedback}.
\end{proof}

\subsection{Convergence of Observer Estimates}\label{subsec:convergenceOfObserverEstimates}
The scaled error dynamics of the EHGO are written by making the following change of variables
\begin{equation}
    \begin{gathered}
        {\eta}_j^i = \frac{({\chi}_j^i - \hat{{\chi}}_j^i)}{\epsilon^{\varrho_i-j}}, \quad \tilde{\bs{\omega}}_s = \bs{\omega}_s - \hat{\bs{\omega}}_s,
    \end{gathered}
\end{equation}
where $\chi_j^i$ is the $j$-th element of $\bs{\chi}_i$ for $1\leq i \leq 3$ and $1\leq j \leq \varrho_i$, and ${\hat{\chi}_j^i}$ is the estimate of $\chi_j^i$ obtained using the EHGO. In the new variables, the scaled EHGO estimation error dynamics become
\begin{equation}\label{eq:fullObserverErrorDynamics}
    \begin{split}
        \epsilon\dot{\bs{\eta}}^i &= F_i\bs{\eta}^i + B_1^i\left[\Delta \bar{f}^i + \bar{G}^i(\bs{\vartheta}_1)\tilde{\bs{\omega}}_s\right]
         + \epsilon B_2^i\varphi^i(t,\bs{q},\bs{p}_c),
    \end{split}
\end{equation}
where
\begin{subequations}
\begin{equation}
    F_{i}=\left[\begin{smallmatrix}{-\alpha_{1}^{i}}I_3 & {I_3} & {\cdots} & {0_3} \\ {\vdots} & {} & {\ddots} & {\vdots} \\ {-\alpha^i_{\varrho_i-1}}I_3 & {0_3} & {\cdots} & {I_3} \\ {-\alpha^i_{\varrho_i}}I_3 & {0_3} & {\cdots} & {0_3}\end{smallmatrix}\right], \quad B_1^i = \left[\begin{smallmatrix} 0_3 \\ \vdots \\ I_3 \\ 0_3 \end{smallmatrix}\right],
\end{equation}
\begin{equation}
    B_2^i = \left[0_3 \ \cdots \ 0_3 \ I_3\right]^\top, \quad \bs{\eta}^i = \left[{\bs{\eta}_1^i}^\top \ \cdots \ {\bs{\eta}_{\varrho_i}^i}^\top\right]^\top,
\end{equation}
\end{subequations}
and $\Delta \bar{f}^i = \bar{f}^i(\bs{\xi},\bs{p}_{c_3},\bs{\vartheta}_1,\dot{\bar{\bs{\vartheta}}}_r) - \bar{f}^i(\hat{\bs{\xi}},\hat{\bs{p}}_{c_3},\bs{\vartheta}_1,\dot{\bar{\bs{\vartheta}}}_r)$ and $\bar{f}^i$, $\bar{G}^i$, and $\varphi^i$ correspond to rows $3i-2$ to $3i$ of $\bar{f}$, $\bar{G}$, and $\varphi$, respectively. Note \eqref{eq:fullObserverErrorDynamics} is an $O(\epsilon)$ perturbation of
\begin{equation}\label{eq:simpleEHGOSystem}
    \epsilon\dot{\bs{\eta}}^i = F_i\bs{\eta}^i + B_1^i\left[\Delta \bar{f}^i + \bar{G}^i(\bs{\vartheta}_1)\tilde{\bs{\omega}}_s\right].
\end{equation}

The actuator error dynamics in terms of the error in squared rotor angular rate, $\tilde{\bs{\omega}}_s$, and rotor angular rate error, $\tilde{\bs{\omega}} = \bs{\omega} - \hat{\bs{\omega}}$ can be written as
\begin{equation}\label{eq:actuatorErrorDynamics}
    \begin{split}
        \tau_m\dot{\tilde{\bs{\omega}}}_s &= -2\tilde{\bs{\omega}}_s + 2W_\text{des}\tilde{\bs{\omega}} \\
        \tau_m\dot{\tilde{\bs{\omega}}} &= -\tilde{\bs{\omega}},
    \end{split}
\end{equation}
where 
$W_\text{des} = \text{diag}[\sqrt{\omega_{s d_i}}] \inr^{n\times n}$ for $i\in\{1,\dots,n\}$ is time-varying. By exploiting the fact that $W_\text{des}$ is bounded, i.e., $\sqrt{\omega_{sd_i}} \leq \omega_{\max}$, for each $i$, where $\omega_{\max} \inr$ is the maximum achievable rotor angular rate, the actuator error dynamics \eqref{eq:actuatorErrorDynamics} can be analyzed as a cascaded system with the Lyapunov functions
\begin{equation}\label{eq:actuatorIndividualLyapunov}
    V_{\tilde{\omega}_s} = \tilde{\bs{\omega}}_s^\top\tilde{\bs{\omega}}_s, \quad V_{\tilde{\omega}} = \tilde{\bs{\omega}}^\top\tilde{\bs{\omega}},
\end{equation}
and the composite Lyapunov function
\begin{equation}\label{eq:v_omega}
    V_\omega = d_2V_{\tilde{\omega}_s} + V_{\tilde{\omega}}, \quad d_2>0,
\end{equation}
where $d_2$ is sufficiently small (see Appendix \ref{A:cascade} for details). Define the set $\Omega_\omega = \{V_\omega \leq c_\omega\}$ where $c_\omega\inr_{>0}$ is an arbitrary constant.

\begin{lemma}[Stability of Actuator Dynamics]\label{lem:actuatorDynamics} 
    For bounded input, $\sqrt{\bs{\omega}_{sd_i}}$ for $i\in\{1,\dots,n\}$, the actuator error dynamics \eqref{eq:actuatorErrorDynamics} will globally exponentially converge to the origin. Therefore, the simulated rotor angular rates, $\hat{\bs{\omega}}$, exponentially converge to the actual rotor angular rates, $\bs{\omega}$.
\end{lemma}
\begin{proof}
    See Appendix \ref{A:proofActuatorDynamics}.
\end{proof}

The systems \eqref{eq:simpleEHGOSystem} and \eqref{eq:actuatorErrorDynamics} form the cascaded system
\begin{equation}\label{eq:EHGOcascadeSystem}
    \begin{split}
        \epsilon\dot{\bs{\eta}}^i &= F_i\bs{\eta}^i + B_1^i\left[\Delta \bar{f}^i + \bar{G}^i(\bs{\vartheta}_1)\tilde{\bs{\omega}}_s\right] \\
        \tau_m\dot{\tilde{\bs{\omega}}}_s &= -2\tilde{\bs{\omega}}_s + 2W_\text{des}\tilde{\bs{\omega}} \\
        \tau_m\dot{\tilde{\bs{\omega}}} &= -\tilde{\bs{\omega}}.
    \end{split}
\end{equation}
We now define the state vector of scaled observer error and actuator error as $\bs{\Delta} = [\bs{\eta}^1\; \bs{\eta}^2\; \bs{\eta}^3 \; \tilde{\bs{\omega}}_s \; \tilde{\bs{\omega}}]^\top$. In comparison with a standard EHGO, \eqref{eq:EHGOcascadeSystem} has additional vanishing perturbation terms with associated dynamics. In the following theorem, we establish that these perturbation terms do not affect the convergence of the EHGO. Furthermore, the perturbation term in \eqref{eq:fullObserverErrorDynamics} is continuous and can be bounded by $\epsilon\norm{\varphi(t,\bs{q},\bs{p}_c)} \leq \epsilon\kappa$ for $\kappa \inr_{>0}$, and can be treated as a nonvanishing perturbation. Using \cite[\textit{Lemma 9.2}]{khalil2002nonlinear}, it can be shown that the perturbed observer error dynamics converge to an $O(\epsilon\kappa)$ neighborhood of the origin.

A Lyapunov function for the EHGO error system \eqref{eq:simpleEHGOSystem} with the input, $\tilde{\bs{\omega}}_s$, set to zero is
\begin{equation}\label{eq:observerErrorLyapunov}
     \epsilon V_\eta = \sum_{i=1}^3(\bs{\eta}^i)^\top P_\eta^i\bs{\eta}^i, \quad P_\eta^i F_i + F_i^\top P_\eta^i = -I_{3\varrho_i}.
\end{equation}
A composite Lyapunov function for \eqref{eq:EHGOcascadeSystem} is
\begin{equation}\label{eq:observerActuatorCompositeLyapunov}
    V_\Delta = d_3V_\eta + V_\omega, \quad d_3>0,
\end{equation}
where $d_3$ is sufficiently small (see Appendix \ref{A:cascade} for details). 

Recall that $(\chi_1^2, \chi_2^2)= (\xi_1, \xi_2)$. Also, the estimates of $(\xi_1, \xi_2)$ can be expressed as 
$\hat \xi_1 = \xi_1 - \epsilon^2 \eta^2_1$ and $\hat \xi_2 = \xi_2 - \epsilon \eta^2_2.$ Consider a strict subset of $\Omega_\xi$, defined by $\supscr{\Omega}{sub}_\xi \subset \Omega_\xi$. Define
\begin{equation}
\Omega_\eta = \setdef{(\bs \eta^1, \bs \eta^2, \bs \eta^3) \in \real^{10}}{(\hat \xi_1, \hat \xi_2)\in \Omega_\xi, \; \forall \bs \xi \in \supscr{\Omega}{sub}_\xi}. 
\end{equation}

Let $c_\Delta \inr_{>0}$ be the largest constant such that $\Omega_\Delta = \{V_\Delta \le c_\Delta\}$ is contained in $\Omega_\eta \times \Omega_\omega$.

\begin{theorem}[Convergence of EHGO Estimates]\label{thm:EHGOConvergence}
    There exists sufficiently small $\epsilon^*$ such that for all $\epsilon \in (0, \epsilon^*)$, $\Omega_\Delta$ is positively invariant, and for each $\bs{\Delta}(0) \in \Omega_\Delta$, $\bs{\Delta}(t)$ converges exponentially to an $O(\epsilon\kappa)$ neighborhood of the origin.
\end{theorem}

\begin{proof}
    See Appendix \ref{A:proofEHGOEstimates}.
\end{proof}

\subsection{Stability Under Output Feedback}
The system under output feedback is a singularly perturbed system that can be split into two timescales. The multi-rotor dynamics and control reside in the slow timescale while the observer and actuator dynamics reside in the fast timescale. We now establish the stability of the overall output feedback system.

\begin{theorem}[Stability Under Output Feedback]\label{thm:OFStability}
    For the output feedback system defined by \eqref{eq:rotationalErrorDynamicsFinal}, \eqref{eq:translationalErrorDynamics}, \eqref{eq:simulatedActuatorDynamics}, \eqref{eq:fullObserverDynamics}, \eqref{eq:rotationalOutputFeedback}, \eqref{eq:translationalOutputFeedback}, and \eqref{eq:omegaSD}, satisfying \textit{Assumptions~\ref{as:disturbance},~\ref{as:disturbanceDynamics}, and~\ref{as:rotationalRefSet}}, the following statements hold
\begin{enumerate}
    \item given any compact subset $\Omega_A \subset \Omega_q \times \Omega_\Delta$, there exists a sufficiently small $\epsilon^*$ such that for any $\epsilon \in (0, \epsilon^*)$, $\Omega_A$ is a positively invariant set;
    \item for $\epsilon \in (0,\epsilon^*)$ the trajectories of the output feedback system  exponentially converge to an $O(\epsilon\kappa)$ neighborhood of the origin with $\Omega_A$ as a subset of its region of attraction. 
\end{enumerate}
\end{theorem}

\begin{proof}
    See Appendix \ref{A:proofOFStability}.
\end{proof}

\section{Numerical Simulation}\label{sec:simulation}
The proposed method is simulated with the reference system taken as a moving ground vehicle on which the multi-rotor will land. However, since the multi-rotor may initially be far from the ground vehicle, i.e., $\bs{p}_1 - \bs{x}_{c_1}$ may be large, we will bound the estimate of this error to prevent overly aggressive maneuvers by saturating $\hat{\bs{\rho}}_1$ as
\begin{equation}\label{eq:boundRho1}
    \hat{\bs{\rho}}_{1_s} = \delta_\rho\operatorname{tanh}(\hat{\bs{\rho}}_1/\delta_p),
\end{equation}
where $\delta_\rho\inr$ is chosen to determine the rate of convergence of the multi-rotor position, $\bs{p}_1$, and the ground vehicle position, $\bs{x}_{c_1}$. The saturated estimate is then used in the output feedback control \eqref{eq:translationalOutputFeedback}. 

\begin{figure}
    \centering
    \includegraphics[width=0.47\textwidth]{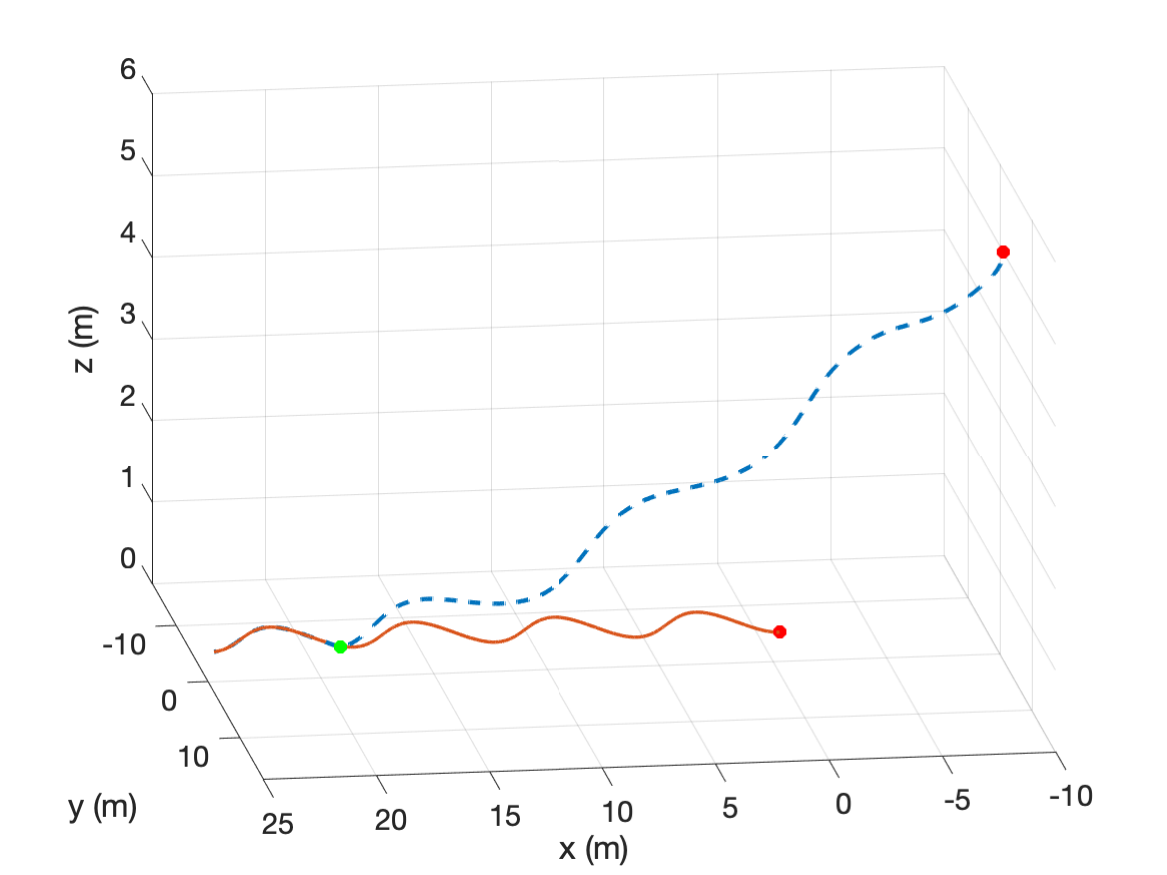}
    \caption{The trajectory of the multi-rotor UAV (dashed) and the trajectory of the ground vehicle (solid). The red points are the initial conditions and the green point signifies the occurrence of the landing.}
    \label{fig:simulationResults}
\end{figure}

\begin{figure}
    \centering
    \includegraphics[width=0.47\textwidth]{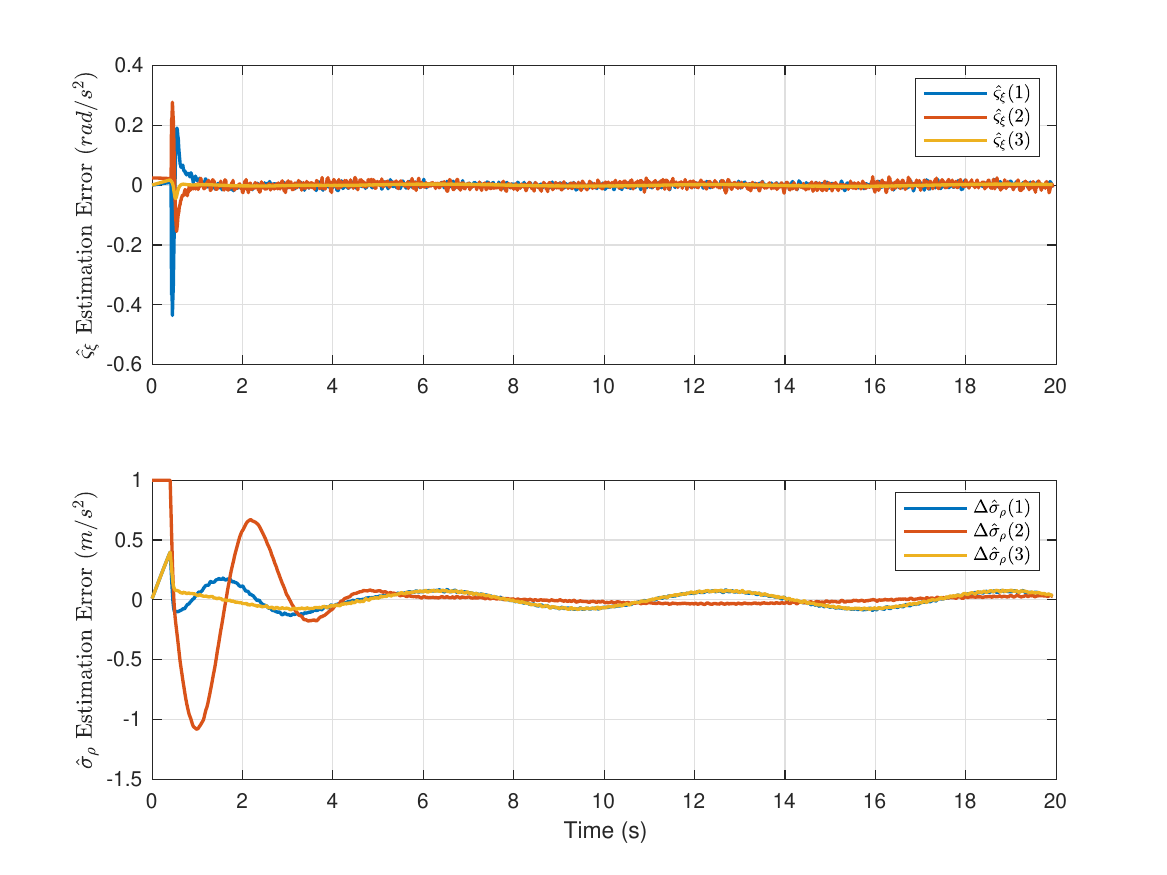}
    \caption{Estimation errors of both rotational disturbance and translational disturbance during the simulation.}
    \label{fig:simEstimationErrors}
\end{figure}

The multi-rotor initial position is $\bs{p}_1(0) = [-10, \ 1, \ -5]^\top$ and the initial position of the ground vehicle is $\bs{p}_{c_1}(0) = [2, \ 0, \ -0.5]^\top$. The ground vehicle follows the trajectory $\bs{p}_{c_1}(t) = [t+2, \ 2\cos(t), \ -0.5]^\top$. While only having a position measurement of the ground vehicle, with added noise, the multi-rotor is able to track and land on the vehicle, as shown in Fig. \ref{fig:simulationResults}. The multi-rotor is able to make this landing while canceling disturbances in both the rotational and translational subsystems, $\bs{\sigma}_\xi = [\sin(t) \ \cos(t) \ \sin(t)]^\top$ and $\bs{\sigma}_\rho = [\cos(t) \ \sin(t) \ \cos(t)]^\top$, respectively. Gaussian white noise is added to all measurement signals. To showcase the ability of the observer to accurately estimate the disturbances which are applied to the system in the simulation, and thus known, the disturbance estimation error for both rotational and translational subsystems are shown in Fig. \ref{fig:simEstimationErrors}.

\section{Experimental Validation}\label{sec:experiment}
The proposed estimation and control method is implemented on an experimental platform to validate performance and show the practical application of this control methodology to landing a multi-rotor on a small moving ground vehicle. 

\begin{figure}
    \centering
    \includegraphics[width=0.45\textwidth]{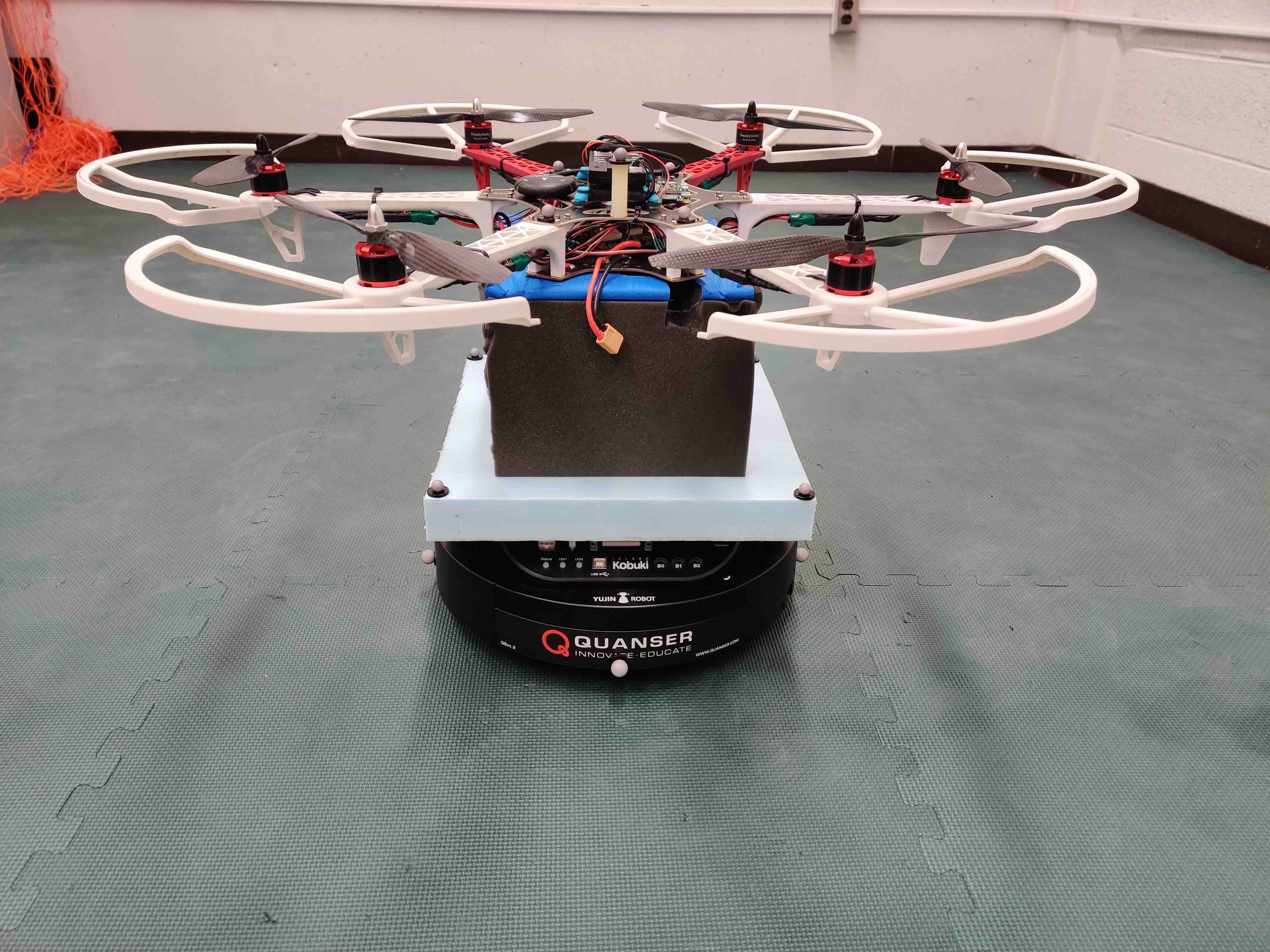}
    \caption{Experimental multi-rotor on ground vehicle landing platform.}
    \label{fig:quadAndGroundVehicle}
\end{figure}

\begin{figure}
    \centering
    \includegraphics[width=0.47\textwidth]{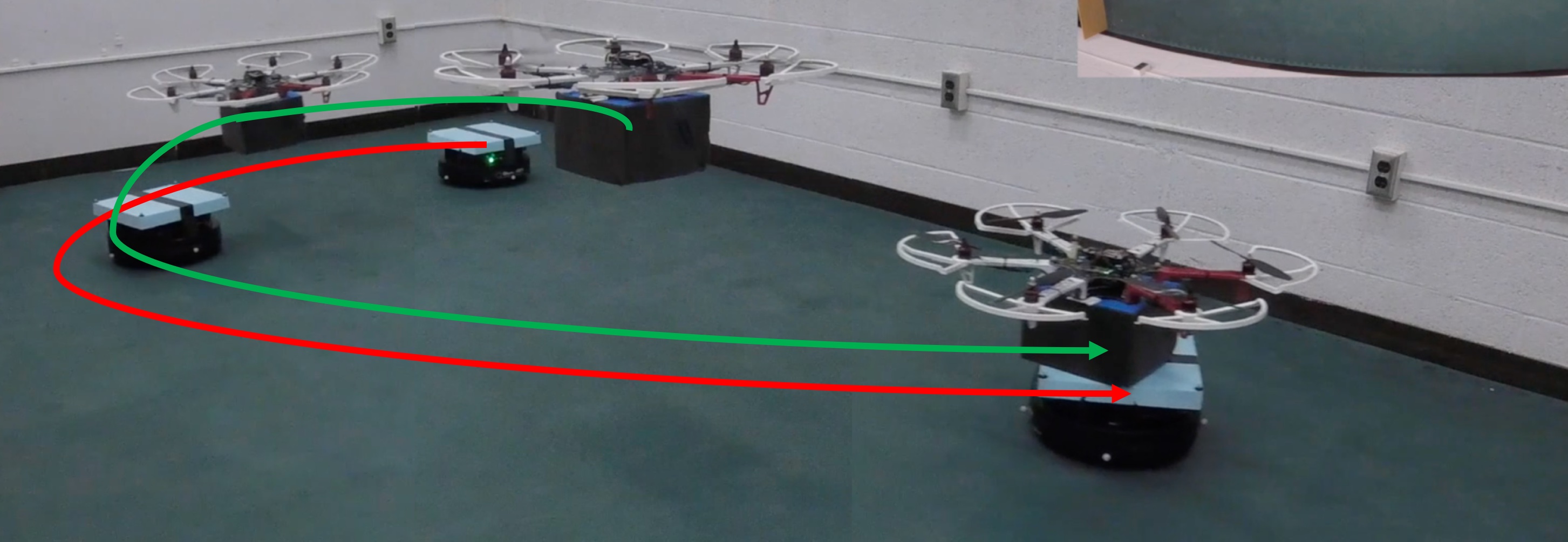}
    \caption{Experimental landing on moving ground vehicle.}
    \label{fig:experimentalLanding}
\end{figure}

\begin{figure}
    \centering
    \includegraphics[width=0.47\textwidth]{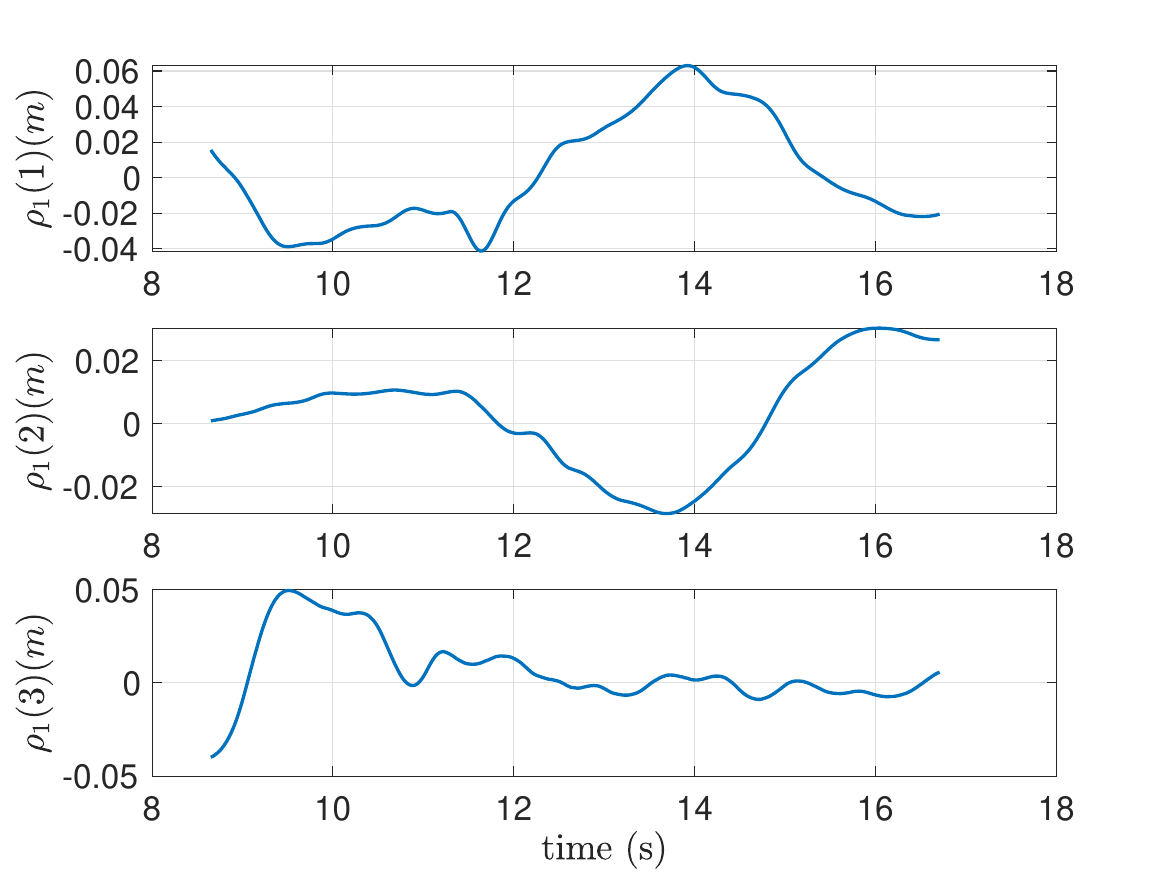}
    \caption{Translational tracking error during flight. }
    \label{fig:translationalTrackingError}
\end{figure}

\begin{figure}
    \centering
    \includegraphics[width=0.47\textwidth]{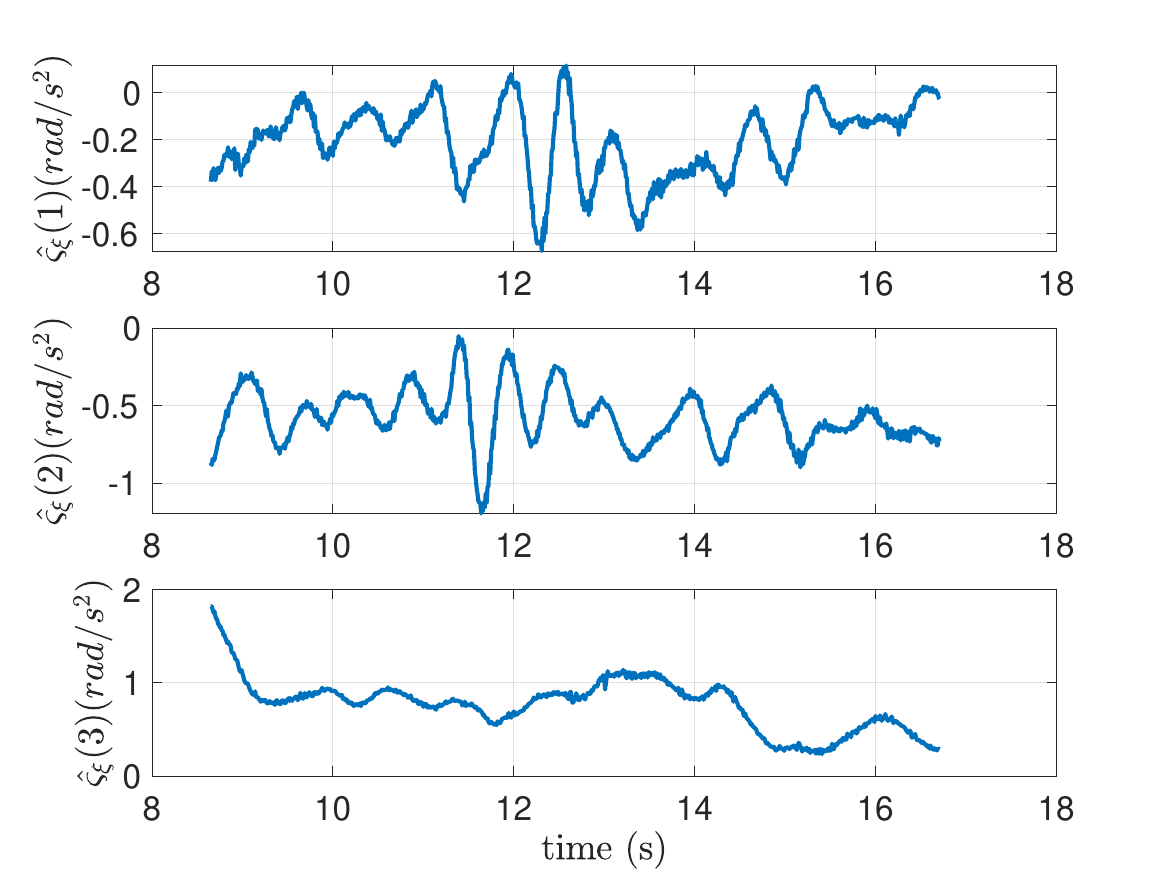}
    \caption{Estimates of the total rotational disturbance affecting the hexrotor during an experimental flight.}
    \label{fig:varsigmaXi}
\end{figure}

\begin{figure}
    \centering
    \includegraphics[width=0.47\textwidth]{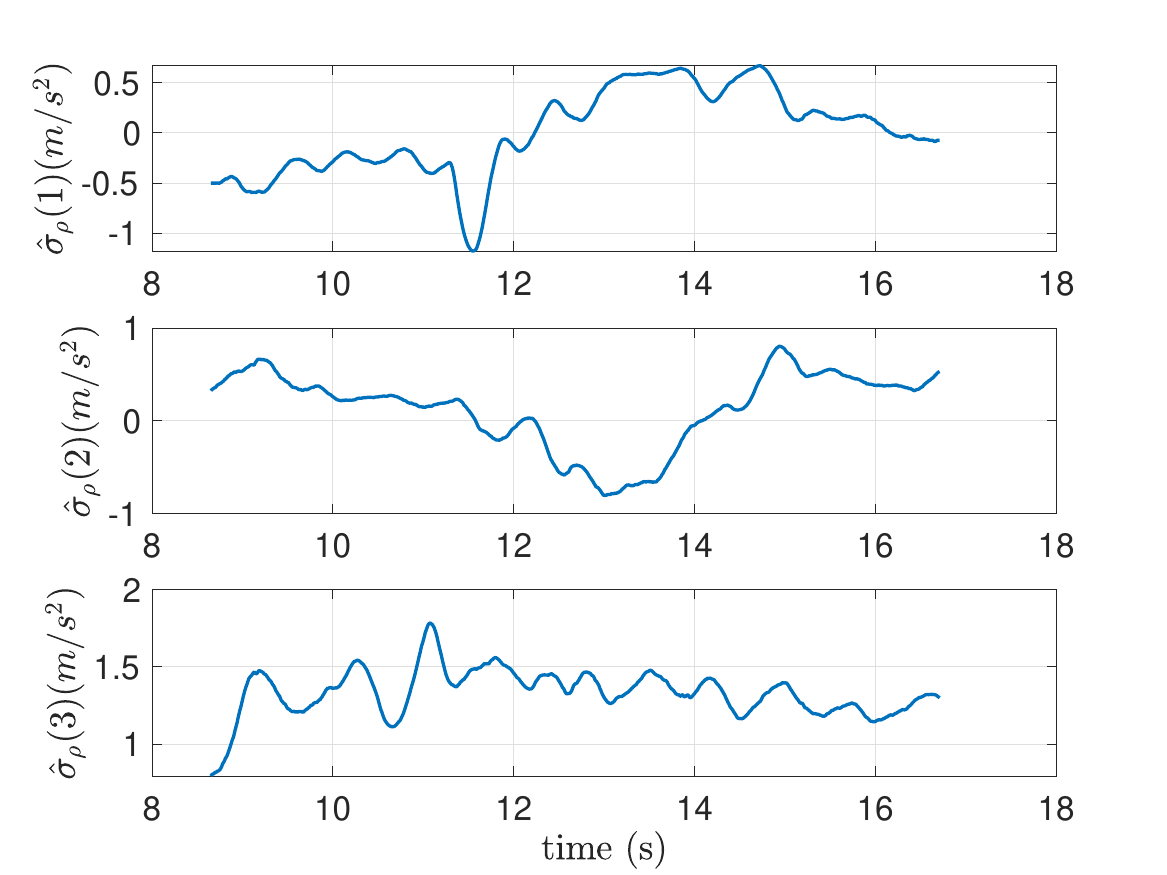}
    \caption{Estimates of the total translational disturbance affecting the hexrotor during an experimental flight.}
    \label{fig:sigmaRho}
\end{figure}

\subsection{Hardware}
The experimental multi-rotor platform is built on a 550mm hexrotor frame with 920kV motors and 10x4.5 carbon fiber rotors. Six 30A electronic speed controllers (ESCs) are used for motor control and the system is powered by a 5000mAh 4s LiPo battery. The model parameters for the experimental platform were found to be
\begin{equation}
    \begin{gathered}
        J = \text{diag}(0.0228 \; 0.0241 \; 0.0446), \quad m = 1.824kg, \\
        M = \left[\begin{smallmatrix} 1 & 1 & 1 & 1 & 1 & 1 \\ -\frac{r}{2} & -r & -\frac{r}{2} & \frac{r}{2} & r & \frac{r}{2} \\ \frac{r\sqrt{3}}{2} & 0 & -\frac{r\sqrt{3}}{2} & -\frac{r\sqrt{3}}{2} & 0 & \frac{r\sqrt{3}}{2} \\ C_D & -C_D & C_D & -C_D & C_D & -C_D \end{smallmatrix}\right], \quad C_D = 0.1, \\
        \quad C_T = 1.8182e-05, \quad r = 0.275m, \quad \tau_m = 0.059.
    \end{gathered}
\end{equation}
The moment of inertia matrix, $J$, was measured using the bifilar pendulum approach \cite{jardin2007optimized}. The mapping matrix, $M$, is derived from the geometry of the airframe, in this case a hexrotor with \texttt{X} geometry with rotors numbered clockwise starting from the front right. The aerodynamic drag of the rotors, $C_D$, and the constant mapping squared actuator speed to force, $C_T$, were obtained using a photo-tachometer to measure rotor angular rate and a load cell to measure the forces generated at a range of speeds. Similarly, the actuator time constant, $\tau_m$, was measured by applying several step inputs of varying magnitude to the rotor, measuring the response with the photo-tachometer, and fitting a first-order system to the data. The length, width, and height of the flying arena are $4.8m$, $3.6m$, and $3.6m$, respectively. Thus, the size of the hexrotor is roughly 1/6-th of the arena, which leads to large aerodynamic effects due to the interaction of rotors' airflow with physical structures. All gains used in the experiments and simulations are shown in Table \ref{tab:params}.

\begin{table}[]
    \centering
    \begin{tabular}{|c|c|}
        \hline
        $[\beta_1, \beta_2]$ = [5, 2] & $[\alpha_1, \alpha_2, \alpha_3, \alpha_4]$ = [10, 6.75, 4.5, 2]\\
        \hline
        $[\gamma_1,\gamma_2]$ = [3, 1.8] & $\epsilon$ = 0.02 \\
        \hline
    \end{tabular}
    \caption{Control Parameters Used in Experiment and Simulation}
    \label{tab:params}
\end{table}

The control method is implemented on a Pixhawk 4 Flight Management Unit (FMU) in discrete time at 100Hz using Mathworks Simulink through the \textit{PX4 Autopilots Support from Embedded Coder} package. This enables the control method to be integrated with the PX4 firmware to run on the Pixhawk 4 hardware. As a result, we can access fused estimates of the vehicle orientation from the EKF running in the PX4 firmware. The position estimates of both the multi-rotor and ground vehicle are pulled from a Vicon server at 100Hz. The estimates are sent over a UDP connection to a Raspberry Pi Zero that is running onboard the multi-rotor. The Raspberry Pi Zero then relays the position information to the FMU over a serial connection.

The ground vehicle is a Quanser QBot2 with a landing platform attached as shown with the multi-rotor on the landing platform in Fig. \ref{fig:quadAndGroundVehicle}. The ground vehicle is manually teleoperated using a joystick through Simulink. This ensures that no prior information about the trajectory is known, as the trajectory is generated in real-time by the operator.

\subsection{Experimental Procedure}

\begingroup
\centering
\begin{figure*}%[t!]
\centering
\includegraphics[width=0.43\textwidth]{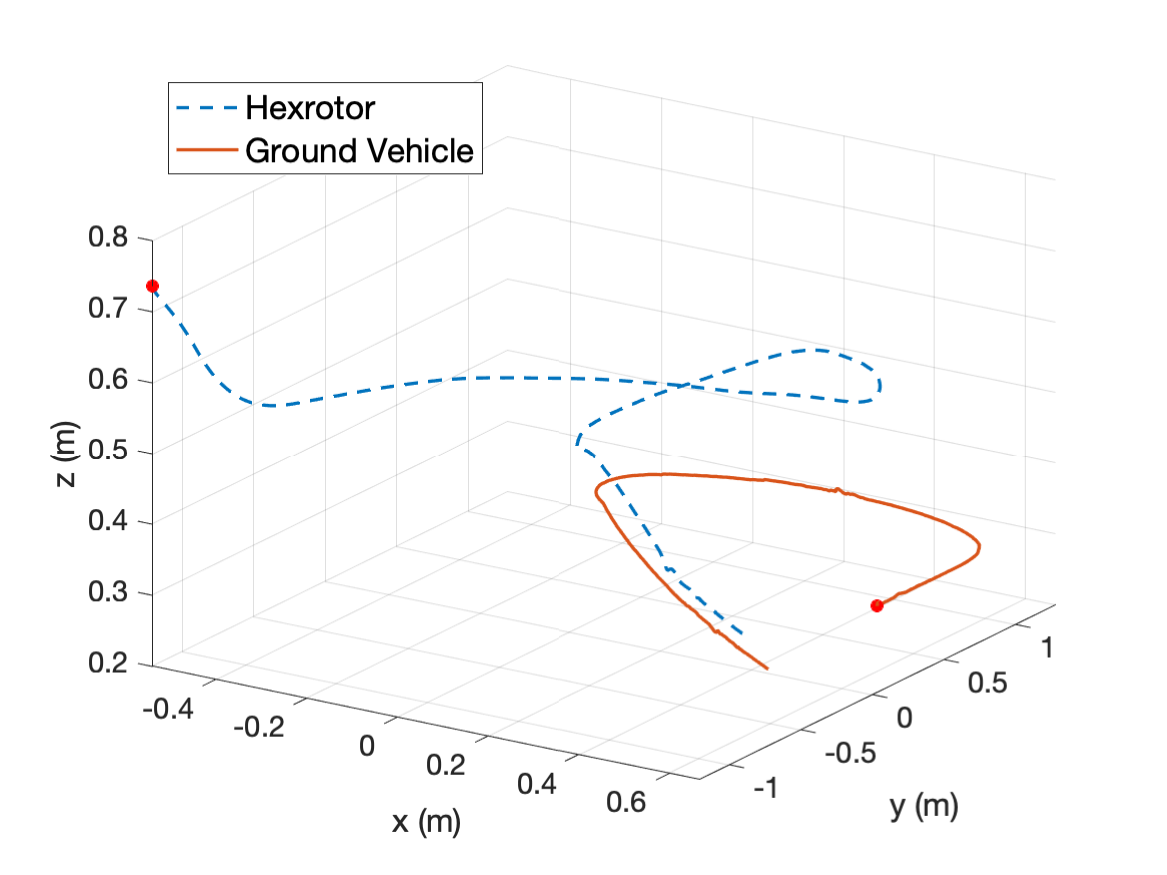}\label{fig:landing1}
\includegraphics[width=0.43\textwidth]{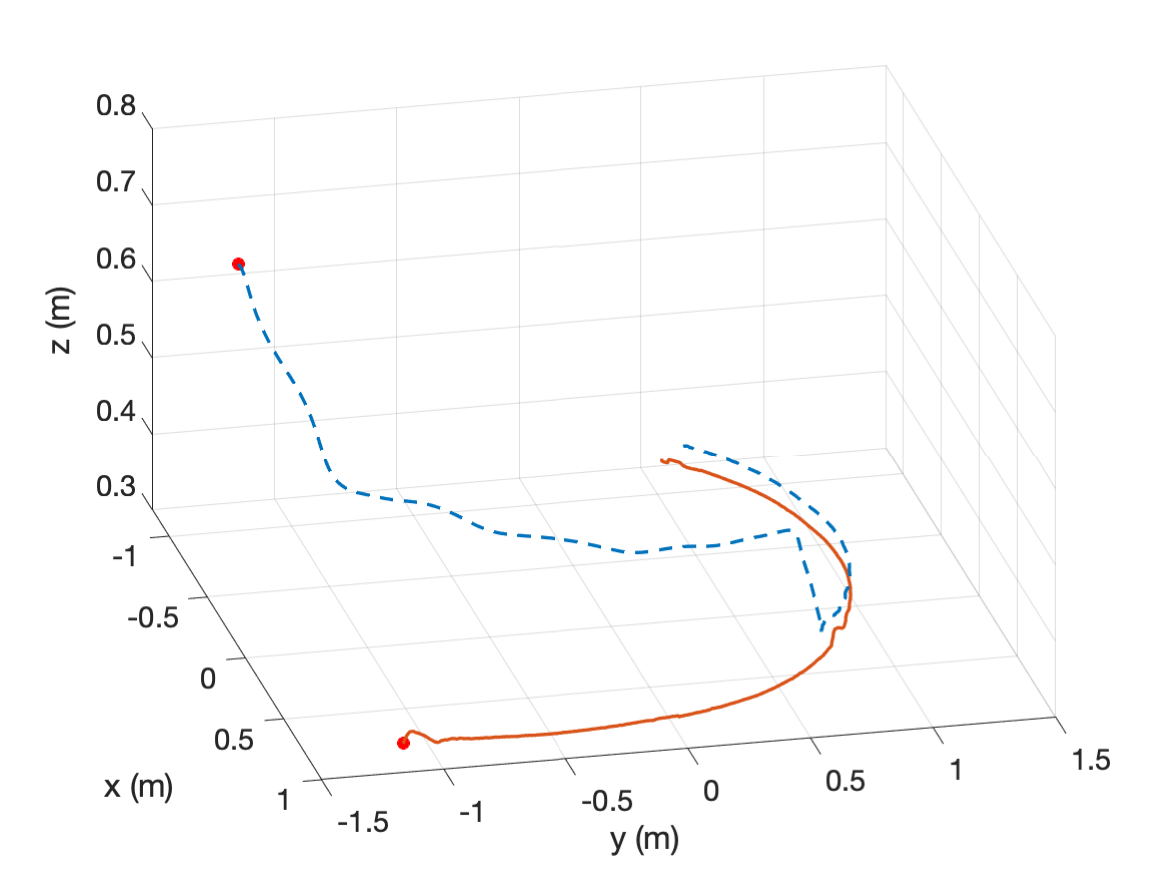}\label{fig:landing2}\\
\includegraphics[width=0.43\textwidth]{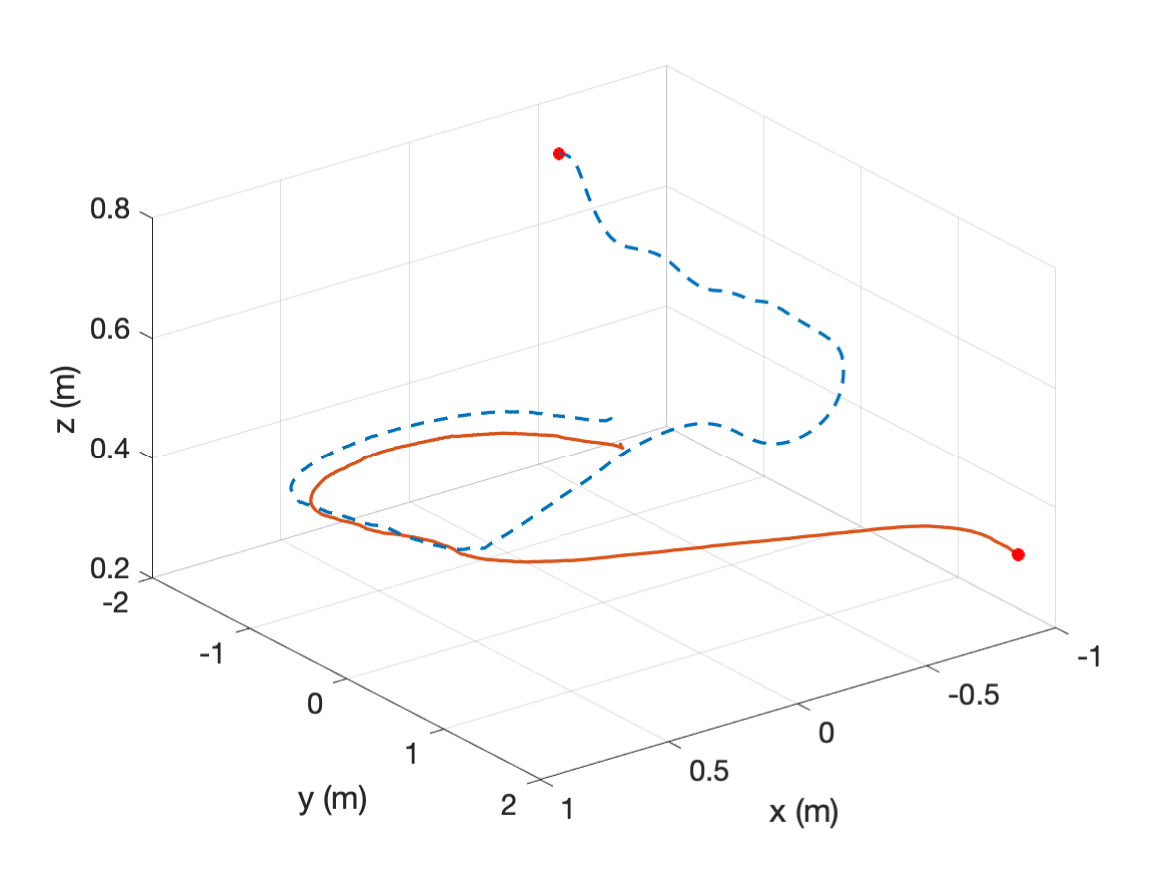}\label{fig:landing3}
\includegraphics[width=0.43\textwidth]{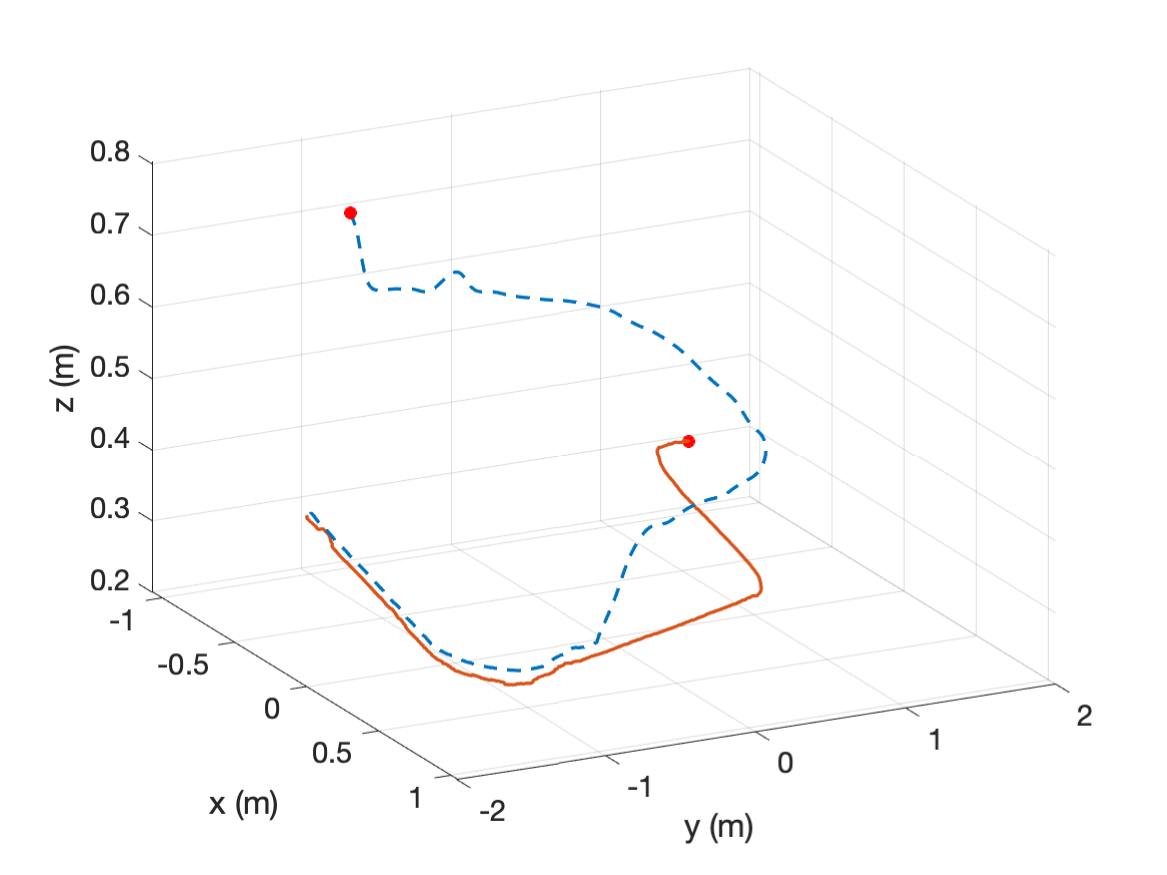}\label{fig:landing4}
\caption{Multiple experimental landing trajectories showing the multi-rotor trajectory (dashed) and the ground vehicle trajectory (solid). The red dots correspond to the initial conditions of the system when a landing was commanded.}
\label{fig:experimentalLandings}
\end{figure*}
\endgroup

The hexrotor initially ascends to a fixed altitude and holds the position until commanded to track and land on the ground vehicle. Once a landing command is sent, the hexrotor begins converging on the position of the ground vehicle while the ground vehicle is being manually teleoperated around the area until the hexrotor successfully lands.

To ensure the large initial position error does not result in overly aggressive control action, the same bounding function \eqref{eq:boundRho1} is used to bound the position error vector $\bs{\rho}_1$. Furthermore, to ensure the multi-rotor approaches the ground vehicle from above, an offset is added to the $z$ component of the reference system. Once the multi-rotor is within some pre-defined radius of the center of the ground vehicle, in this case, $4cm$, the offset is removed so the hexrotor will commence landing on the ground vehicle.

Multiple experimental test flights were conducted with different initial conditions for both the hexrotor and ground vehicle. Each test was also performed with different ground vehicle trajectories. These experiments show the ability of the algorithm to successfully land regardless of differences in initial conditions or different reference trajectories.

The ground vehicle trajectories and the hexrotor trajectories are shown for four different experimental flights in Fig. \ref{fig:experimentalLandings}. The overall translational tracking error, $\bs{\rho}_1$, is shown in Fig. \ref{fig:translationalTrackingError}, with a terminal error at landing of only $[-2, 2.7, 0.5]$cm. The estimates of the disturbances affecting the system in the rotational and translational dynamics for one such flight are shown in Fig. \ref{fig:varsigmaXi} and Fig. \ref{fig:sigmaRho}, respectively, where the first eight seconds are omitted and correspond to the initial takeoff and climb to altitude. Notice that the translational disturbance estimate, specifically $\hat{\bs{\sigma}}_\rho(3)$ in Fig. \ref{fig:sigmaRho}, contains a constant offset. This offset is a result of the charge state of the battery. As the battery voltage decreases, the thrust applied by the rotors for a given commanded speed decreases. Also, large rotational disturbances arise in Fig. \ref{fig:varsigmaXi}, which can be caused by unmodeled aerodynamic effects discussed earlier, inaccuracies in the inertia matrix, or differences between speed controllers. We do not model these discrepancies, however, the observer is able to estimate and compensate for these uncertainties in the control to result in excellent tracking performance. 
A video of the experiments can be found at \url{https://youtu.be/oWcl4ydNLDs}

\section{Conclusions and Future Directions}\label{sec:conclusion}
We studied a real-time trajectory estimation and tracking problem for a multi-rotor in the presence of modeling errors and external disturbances. The unknown trajectory is generated from a dynamical system with unknown or partially known dynamics. We designed and rigorously analyzed an EHGO-based output feedback controller to guarantee the stable operation of the overall system. 

The capability of the controller is illustrated using the example of a multi-rotor landing on a moving ground vehicle. The multi-rotor landing is shown in simulation with noise and disturbances added, as well as implemented experimentally on a hexrotor platform. Multiple initial conditions and unknown trajectories are tested experimentally and shown to result in successful landings.

We plan to extend this method to consider control optimality. The feedback linearizing control could be replaced with an optimal control strategy, such as model predictive control. Furthermore, the estimates of disturbance from the EHGO could be used to parameterize a disturbance model online for use in control design. This work could additionally be extended to a fully self-contained system by implementing vision based techniques for estimating the relative position to the ground vehicle \cite{falanga2017vision,paris2020dynamic}. If measurement noise was found to be a problem on the experimental system, other observer designs could be utilized, for example a low-power EHGO \cite{astolfi2018low} or a cascaded high-gain observer \cite{khalil2017cascade}.

\section*{Acknowledgement}
We would like to thank Professor Hassan K. Khalil for his invaluable insights on extended high-gain observer design and analysis.

\appendix
\section{Proof of Lemma \ref{lem:restrictingDomainOfAnalysis} (Restricting Domain of Operation)}\label{A:proofRestrictingDomain}

Substituting $P_\xi$ in \eqref{eq:v_xi}, the rotational tracking error Lyapunov function can be written as
\begin{equation}
    % \small
    V_\xi = \frac{(\beta_1 + 1)\bs{\xi}_1^\top \bs{\xi}_1}{2\beta_2} + \frac{\beta_1\bs{\xi}_2^\top \bs{\xi}_2 + (\beta_2\bs{\xi}_1 + \bs{\xi}_2)^\top(\beta_2\bs{\xi}_1 + \bs{\xi}_2)}{2\beta_1\beta_2}.
\end{equation}
Taking the bound on the Lyapunov function
\begin{equation}
    V_\xi \leq c_\xi \Rightarrow \frac{(\beta_1 + 1)\bs{\xi}_1^\top\bs{\xi}_1}{2\beta_2} \leq c_\xi,
\end{equation}
and choosing $c_\xi$ in the following manner
\begin{equation}
    c_\xi < \frac{(\beta_1 + 1)\delta^2}{2\beta_2} \Rightarrow \norm{\bs{\xi}_1(t)} < \delta,
\end{equation}
over the set $\Omega_\xi$. Since $P_\xi$ is positive definite, the Lyapunov function \eqref{eq:v_xi} satisfies the following inequalities
\begin{equation}
    \lambda_{\min} (P_\xi)\norm{\bs{\xi}}^2 \leq V_\xi \leq \lambda_{\max} (P_\xi)\norm{\bs{\xi}}^2,
\end{equation}
where $\lambda_\text{min}(\cdot)$ is the minimum eigenvalue of the argument.
Also, using \eqref{eq:rotationalClosedLoopStateFeedback} and \eqref{eq:v_xi}, we have $\dot{V}_\xi = \bs{\xi}^\top ( A_\xi^\top P_\xi + P_\xi A_\xi ) \bs{\xi} $ and with \eqref{eq:v_xi_bnd} it yields
\begin{equation}
    \dot{V}_\xi = -\bs{\xi}^\top \bs{\xi} \leq -\norm{\bs{\xi}}^2,
\end{equation}
 showing that $\Omega_\xi$ is positively invariant.

In view of potential rotational tracking errors that show up in \eqref{eq:translationalClosedLoopStateFeedback} as $e_\vartheta(t,\bs{\xi}_1)$, the translational tracking error Lyapunov function \eqref{eq:v_rho} satisfies the following inequalities when substituting in the dynamics \eqref{eq:translationalClosedLoopStateFeedback}
\begin{equation}
    \begin{gathered}
        \lambda_{\min} (P_\rho)\norm{\bs{\rho}}^2 \leq V_\rho \leq \lambda_{\max} (P_\rho)\norm{\bs{\rho}}^2,\\
  \text{and} \quad       \dot{V}_\rho \leq -\norm{\bs{\rho}}^2 + 2[0_{3\times 1} \; e_\vartheta(t,\bs{\xi}_1)^\top]^\top P_\rho \bs{\rho}.
    \end{gathered}
\end{equation}
Since $e_\vartheta(t,\bs{\xi}_1)$ and its partial derivatives are continuous on $\Omega_\xi$, and $e_\vartheta$ is uniformly bounded in time, $e_\vartheta$ is Lipschitz in $\bs{\xi}_1$ on $\Omega_\xi$. We can now define
\begin{equation}
    \norm{e_\vartheta(t,\bs{\xi}_1) - e_\vartheta(t,0)} \leq L_e\norm{\bs{\xi}_1} \leq L_e \delta,
\end{equation}
for the Lipschitz constant, $L_e$. We can then bound the translational Lyapunov function derivative by
\begin{equation}
    \dot{V}_\rho \leq -\norm{\bs{\rho}}^2 + 2 L_e \delta \lambda_\text{max}(P_\rho)\norm{\bs{\rho}}.
\end{equation}
For $\norm{\bs{\rho}} > 2 L_e \delta \lambda_\text{max}(P_\rho)$, $\dot{V}_\rho < 0$. Since $V_\rho \leq \lambda_\text{max}(P_\rho)\norm{\bs{\rho}}^2$ we can choose
\begin{equation}
    c_\rho > \lambda_\text{max}(P_\rho)(2 L_e \delta \lambda_\text{max}(P_\rho))^2.
\end{equation}
By this choice, $\dot{V}_\rho < 0$ for $V_\rho \geq c_\rho$, hence $\Omega_\rho$ is compact and positively invariant. Thus, the domain of operation $\Omega_q = \Omega_\xi \times \Omega_\rho$ is positively invariant. $\hfill\blacksquare$

\section{Stability of Generalized Cascade Systems}\label{A:cascade}
A generalized stability proof for cascade systems is adapted from Appendix C.1 of \cite{khalil2015nonlinear}. Consider the cascade connection of two systems
\begin{equation}\label{eq:genericCascadedSystem}
    \dot{\eta} = f_1(t,\eta,\xi), \quad \dot{\xi} = f_2(\xi),
\end{equation}
where $f_1$ and $f_2$ are locally Lipschitz and $f_1(t,0,0)=0$, $f_2(0)=0$. Assuming the origin of $\dot{\xi} = f_2(\xi)$ is exponentially stable, there is a continuously differentiable Lyapunov function, $V_2(\xi)$, that satisfies the following inequalities
\begin{subequations}\label{eq:generalXiLyapunovInequalities}
\begin{equation}
    c_1\norm{\xi}^2 \leq V_2(\xi) \leq c_2\norm{\xi}^2
\end{equation}
\begin{equation}\label{eq:V4Inequalities}
    \quad \frac{\partial V_2(\xi)}{\partial \xi}f_2(\xi) \leq -c_3\norm{\xi}^2
\end{equation}
\begin{equation}
    \norm{\frac{\partial V_2(\xi)}{\partial \xi}} \leq c_4\norm{\xi},
\end{equation}
\end{subequations}
over the set $\Omega_2 = \{V_2(\xi) < c_5\}$ for some $c_5 \inr_{>0}$.

Now, suppose there is a continuously differentiable Lyapunov function, $V_1(\eta)$, that satisfies the inequalities
\begin{equation}\label{eq:V1Inequalities}
    \frac{\partial V_1(\eta)}{\partial \eta}f_1(t,\eta,0) \leq -c\norm{\eta}^2, \quad \norm{\frac{\partial V_1(\eta)}{\partial \eta}} \leq k\norm{\eta},
\end{equation}
over the set $\Omega_1 = \{V_1(\eta) < c_6\}$ for some $c_6 \inr_{>0}$.

Take a composite Lyapunov function for the cascaded system as
\begin{equation}
    V(\eta,\xi) = b V_1(\eta) + V_2(\xi), \quad b > 0,
\end{equation}
in which $b$ can be arbitrarily chosen. The derivative, $\dot{V}$, satisfies
\begin{equation}
    \begin{split}
        \dot{V}(\eta,\xi) &= b\frac{\partial V_1(\eta)}{\partial \eta}f_1(t,\eta,0) + \\
        &b\frac{\partial V_1(\eta)}{\partial \eta}\left[f_1(t,\eta,\xi) - f_1(t,\eta,0)\right] + \frac{\partial V_2(\xi)}{\partial \xi}f_2(\xi)\\
        \dot{V}(\eta,\xi) &\leq -b c\norm{\eta}^2 + b k L\norm{\eta}\norm{\xi} - c_3\norm{\xi}^2,
    \end{split}
\end{equation}
where $f_1$ is Lipschitz in $\xi$ on $\Omega_2$, and $L$ is the associated Lipschitz constant.

The inequality can be written in a quadratic form as
\begin{equation}
    \begin{split}
        \dot{V}(\eta,\xi) &\leq -\left[\begin{matrix} \norm{\eta} \\ \norm{\xi}\end{matrix}\right]^\top \left[\begin{matrix} b c & \frac{-b k L}{2}\\ \frac{-b k L}{2} & c_3 \end{matrix}\right] \left[\begin{matrix} \norm{\eta} \\ \norm{\xi}\end{matrix}\right] \\
        &= -\left[\begin{matrix}\norm{\eta} \\ \norm{\xi}\end{matrix}\right]^\top Q \left[\begin{matrix} \norm{\eta} \\ \norm{\xi}\end{matrix}\right] \leq -\lambda_{\min}(Q)\norm{\left[\begin{matrix} \norm{\eta} \\ \norm{\xi}\end{matrix}\right]}^2,
    \end{split}
\end{equation}
where $b$ is chosen such that $b < 4cc_3/(k L)^2$ to ensure $Q$ is positive definite. The foregoing analysis shows that the origin of \eqref{eq:genericCascadedSystem} is exponentially stable on the set $\Omega = \Omega_1 \times \Omega_2$.

\section{Proof of Theorem \ref{thm:SFStability} (Stability Under State Feedback)}\label{A:proofStateFeedback}
The translational and rotational closed-loop systems can be written as a cascaded system in the following form
\begin{equation}\label{eq:cascadedSystem}
    \begin{aligned}[c]
        \dot{\bs{\rho}}_1 &= \bs{\rho}_2 \\
        \dot{\bs{\rho}}_2 &= -\gamma_1\bs{\rho}_1 - \gamma_2\bs{\rho}_2 + \bs{e}_\vartheta(t,\bs{\xi}_1)\\
        \dot{\bs{\xi}}_1 &= \bs{\xi}_2\\
        \dot{\bs{\xi}}_2 &= -\beta_1\bs{\xi}_1 - \beta_2\bs{\xi}_2
    \end{aligned}
    \ \Rightarrow \
    \begin{aligned}
        \dot{\bs{\rho}}_1 &= \bs{\rho}_2\\
        \dot{\bs{\rho}}_2 &= f_1(t,\bs{\rho},\bs{\xi})\\
        \dot{\bs{\xi}}_1 &= \bs{\xi}_2\\
        \dot{\bs{\xi}}_2 &= f_2(\bs{\xi}).
    \end{aligned}
\end{equation}
Taking the Lyapunov functions for the rotational and translational subsystems, \eqref{eq:v_xi} and \eqref{eq:v_rho}, a composite Lyapunov function can be written
\begin{equation}
    V_{sf} = d_1V_\rho + V_\xi, \quad d_1>0.
\end{equation}
Since $V_\rho$ satisfies \eqref{eq:V1Inequalities} on $\Omega_\rho$, $V_\xi$ satisfies \eqref{eq:generalXiLyapunovInequalities} on $\Omega_\xi$, and $f_1(t,\bs{\rho},\bs{\xi})$ is Lipshitz in $\bs{\xi}$ on $\Omega_\xi$, it can be shown following the generalized proof in Appendix \ref{A:cascade} that for $d_1$ small enough, the entire closed-loop state feedback system converges exponentially to the origin for any trajectory starting within the domain of operation, $\Omega_q$. $\hfill\blacksquare$

\section{Proof of Lemma \ref{lem:actuatorDynamics} (Stability of Actuator Dynamics)}\label{A:proofActuatorDynamics}
The Lyapunov functions for the actuator dynamics \eqref{eq:actuatorErrorDynamics} are $V_{\tilde{\omega}_s}$ and $V_{\tilde{\omega}}$ from \eqref{eq:actuatorIndividualLyapunov}, with the composite Lyapunov function \eqref{eq:v_omega}. Since $V_{\tilde{\omega}_s}$ satisfies \eqref{eq:V1Inequalities} globally, $V_{\tilde{\omega}}$ satisfies \eqref{eq:generalXiLyapunovInequalities} globally, and $\dot{\tilde{\bs{\omega}}}_s$ is globally Lipschitz in $\tilde{\bs{\omega}}$ since $W_\text{des}$ is bounded, using the general result for cascaded systems in Appendix \ref{A:cascade}, it can be shown that the origin is globally exponentially stable when $d_2$ is chosen small enough.  $\hfill\blacksquare$

\section{Proof of Theorem \ref{thm:EHGOConvergence} (Convergence of EHGO Estimates)}\label{A:proofEHGOEstimates}
The Lyapunov function for the actuator error system is \eqref{eq:v_omega} and the Lyapunov function for the EHGO error system with the input, $\tilde{\bs{\omega}}_s$, set to zero is \eqref{eq:observerErrorLyapunov}. A composite Lyapunov function for the cascaded system \eqref{eq:EHGOcascadeSystem} is \eqref{eq:observerActuatorCompositeLyapunov}. The function $\bar{f}^i(\bs{\xi},\bs{x}_{c_3},\bs{\vartheta}_1,\dot{\bar{\bs{\vartheta}}}_r)$ is Lipschitz in $\bs{\xi}$ and  $\bs{x}_{c_3}$ on $\Omega_\xi \times \Omega_{x_c}$ and $(\hat \xi_1, \hat \xi_2) \in \Omega_\xi$. Thus, $\Delta \bar{f}^i$ can be bounded by
\begin{equation}\label{eq:deltafi_bound}
    \norm{\Delta \bar{f}^i} \leq L_\eta\norm{\bs{\chi}_i - \hat{\bs{\chi}}_i} \Rightarrow \norm{\Delta \bar{f}^i} \leq \epsilon L_\eta\norm{\bs{\eta}^i},
\end{equation}
leading to the following bound on the derivative of the Lyapunov function
\begin{equation}
    \begin{split}
        \epsilon \dot{V}_\eta &\leq \sum_{i=1}^3\left(-\norm{\bs{\eta}^i}^2 + 2\epsilon L_\eta\norm{\bs{\eta}^i}^2\norm{P_\eta^i B_1^i}\right) \\
        \epsilon \dot{V}_\eta &\leq -\norm{\bs{\eta}}^2 + 2\epsilon L_\eta\norm{N}\norm{\bs{\eta}}^2,
    \end{split}
\end{equation}
where the elements of the diagonal matrix $N$ are $N_i = \norm{P_\eta^i B_1^i}$. Since $\alpha_j^i$ are tunable and $\epsilon$ is a design parameter, pick $\epsilon$ such that $2\epsilon L_\eta \norm{N} \leq \frac{1}{2}$ resulting in the following inequality
\begin{equation}
    \epsilon \dot{V}_\eta \leq -\frac{1}{2}\norm{\bs{\eta}}^2.
\end{equation}
The composite Lyapunov function \eqref{eq:observerActuatorCompositeLyapunov} consists of $V_\eta$ and $V_\omega$, where $V_\eta$ satisfies \eqref{eq:V1Inequalities} on $\Omega_\eta$, $V_\omega$ satisfies \eqref{eq:generalXiLyapunovInequalities} on $\Omega_\omega$, and $\epsilon\dot{\bs{\eta}}$ is Lipschitz in $\tilde{\bs{\omega}}_s$ on $\Omega_\omega$. Following Appendix \ref{A:cascade}, the origin of \eqref{eq:EHGOcascadeSystem} is exponentially stable for any trajectory starting in $\Omega_\Delta$. Furthermore, the cascade connection of the complete scaled observer error system \eqref{eq:fullObserverErrorDynamics} and the actuator error dynamics \eqref{eq:actuatorErrorDynamics} is the same as \eqref{eq:EHGOcascadeSystem} with perturbation. The perturbation is bounded by $\epsilon\varphi(t,\bs{q},\bs{x}_c)<\epsilon \kappa$ and is continuous, therefore it can be treated as a nonvanishing perturbation. Following \cite[\textit{Lemma 9.2}]{khalil2002nonlinear}, the estimation error of the EHGO converges exponentially to an $O(\epsilon\kappa)$ neighborhood of the origin. Furthermore, $\Omega_\Delta$ will remain invariant under the nonvanishing perturbation. $\hfill\blacksquare$

\section{Proof of Theorem \ref{thm:OFStability} (Stability Under Output Feedback)}\label{A:proofOFStability}
The existence of sufficiently small $\epsilon^*$ such that $\Omega_A$ is invariant can be established analogously to~\cite[\textit{Theorem 14.6}]{khalil2002nonlinear}.
The entire output feedback closed-loop system can now be written in the singularly perturbed form
\begin{subequations}\label{eq:singularlyPerturbedForm}
\begin{align}
        \hspace*{-0.5cm} \dot{\bs{q}} &= A_c\bs{q} + \Delta(\bs{\eta})\label{eq:slowSystem} \\
        \epsilon\dot{\bs{\eta}}^i &= F_i\bs{\eta}^i + B_1^i\left[\Delta \bar{f}^i + \bar{G}^i(\bs{\vartheta}_1)\tilde{\bs{\omega}}_s\right] + \epsilon B_2^i\varphi^i(t,\bs{q},\bs{x}_c) \label{eq:boundaryLayerSystem1} \\
        \tau_m\dot{\tilde{\bs{\omega}}}_s &= -2\tilde{\bs{\omega}}_s + 2W_\text{des}\tilde{\bs{\omega}} \label{eq:boundaryLayerSystem2}\\
        \tau_m\dot{\tilde{\bs{\omega}}} &= -\tilde{\bs{\omega}}, \label{eq:boundaryLayerSystem3}
\end{align}
\end{subequations}
where
\begin{equation}
    A_c = \left[\begin{smallmatrix} A_\rho & 0_6 \\ 0_6 & A_\xi \end{smallmatrix}\right].
\end{equation}
The term $\Delta(\bs{\eta})$ is due to estimation errors and is $O(\epsilon\kappa)$ and can be defined by
\begin{equation}
    \Delta(\bs{\eta}) = \left[\begin{smallmatrix} 0_{3\times1} \\ \gamma_1\epsilon^2\bs{\eta}_1^1 + \gamma_2\epsilon\bs{\eta}_2^1 + \bs{\eta}_3^1 - \bs{\eta}_3^3 \\ 0_{3\times1} \\\beta_1\epsilon^2\bs{\eta}_1^2 + \beta_2\epsilon\bs{\eta}_2^2 + \bs{\eta}_3^2 + \Delta \bar{f}^i\end{smallmatrix}\right].
\end{equation}
where $\norm{\Delta \bar{f}^i} \leq \epsilon L_\eta\norm{\bs{\eta}^i}$ from \eqref{eq:deltafi_bound}.

First, we ignore the last term, $\epsilon B^i \varphi^i(t,\bs{q},\bs{p}_c)$, in the $\bs{\eta}$ dynamics. In this case, the closed-loop system has a two-timescale structure because $\epsilon$ and $\tau_m$ are small. Since the effect of $\Delta(\bs{\eta})$ in \eqref{eq:slowSystem} vanishes as $\epsilon$ is pushed to zero, the boundary layer system can be taken as \eqref{eq:boundaryLayerSystem1}--\eqref{eq:boundaryLayerSystem3} and the slow dynamics can be taken as \eqref{eq:slowSystem}. From \textit{Theorem \ref{thm:EHGOConvergence}}, the origin of the boundary layer system is an exponentially stable equilibrium point as $\epsilon \to 0$, and from \textit{Theorem \ref{thm:SFStability}}, the origin of the slow system is an exponentially stable equilibrium point. 

With the inclusion of $\epsilon B^i \varphi^i(t,\bs{q},\bs{p}_c)$ in the $\bs{\eta}$ dynamics, the overall system is an $O(\epsilon k)$ perturbation of an exponentially stable system. Therefore, similar to \cite[\textit{Lemma 9.2}]{khalil2002nonlinear}, it can be shown that the entire closed-loop system with output feedback control \eqref{eq:singularlyPerturbedForm} will converge to an $O(\epsilon\kappa)$ neighborhood of the origin for any trajectory starting in $\Omega_q \times \Omega_\Delta$. $\hfill\blacksquare$

\section{Peaking Phenomenon}\label{A:peakingPhenomenon}
The EHGO estimation error $\tilde{\chi}_i = \chi_i - \hat{\chi}_i$ can be bounded by
\begin{equation}
    |\tilde{\chi}_i| \leq \frac{b}{\epsilon^{\varrho - 1}}\norm{\bs{\tilde{\chi}}(0)}e^{-at/\epsilon},
\end{equation}
for some positive constants $a$ and $b$, by \textit{Theorem 2.1} in \cite{khalil2017HGO}. Initially, the estimation error can be very large, i.e., $O(1/\epsilon^{\varrho-1})$, but will decay rapidly. To prevent the peaking of the estimates from entering the plant during the initial transient, the output feedback controller needs to be saturated. This is done by saturating the individual estimates outside a compact set of interest using \eqref{eq:estimateSaturation}.

There is some set $\{V_\eta\leq \epsilon^2c\}$ for some $c\inr_{>0}$ that the estimation error will enter after some short time, $T(\epsilon)$, where $\lim_{\epsilon \rightarrow 0}T(\epsilon) = 0$. Since the initial state $\bs{q}(0)$ resides on the interior of the modified compact set of \textit{Theorem \ref{thm:OFStability}}, $\Omega_A$, choosing $\epsilon$ small enough will ensure that $\bs{q}$ will not leave $\Omega_A$ during the interval $[0,T(\epsilon)]$. This establishes the boundedness of all states.

{\footnotesize
\bibliographystyle{asmems4}
\bibliography{references.bib}
}

\end{document}